\documentclass[manuscript]{aastex}

\usepackage{graphicx}
\usepackage{txfonts}
\usepackage{rotating}
\usepackage{pdflscape}
\usepackage{longtable}
\usepackage{gensymb}

\newcommand{\myemail}{mfilho@astro.up.pt}

\shorttitle{The KS Relation in XMPs}
\shortauthors{Filho et al.}

\begin{document}


\title{The Kennicutt-Schmidt Relation in Extremely Metal-Poor Dwarf Galaxies}


\author{M. E. Filho\altaffilmark{1,2,3,4,5}\email{\myemail}, J. S\' anchez Almeida\altaffilmark{1,2}, R. Amor\'\i n\altaffilmark{6}, C. Mu\~noz-Tu\~n\'on\altaffilmark{1,2}, B. G. Elmegreen\altaffilmark{7} and D. M. Elmegreen\altaffilmark{8}} 


\altaffiltext{1}{Instituto Astrof\'\i sica de Canarias, 38200 La Laguna, Tenerife, Spain}
\altaffiltext{2}{Departamento de Astrof\'\i sica, Universidad La Laguna, 38206 La Laguna, Tenerife, Spain}
\altaffiltext{3}{Centro de Astrof\'isica da Universidade do Porto, Rua das Estrelas s/n, 4150-762 Porto, Portugal}
\altaffiltext{4}{Instituto de Astrof\'\i sica e Ci\^encias do Espa\c co,  Universidade de Lisboa, OAL, Tapada da Ajuda, 1349-018 Lisboa, Portugal}
\altaffiltext{5}{SIM/CENTRA, Portugal}
\altaffiltext{6}{National Institute for Astrophysics, Astronomical Observatory of Rome, Via Frascati 33, I-00040 Monteporzio Catone (Rome), Italy}
\altaffiltext{7}{IBM, T. J. Watson Research Center, 1101 Kitchawan Road, Yorktown Heights, NY 10598, USA}
\altaffiltext{8}{Department of Physics and Astronomy, Vassar College, Poughkeepsie, NY 12604, USA}



\begin{abstract}

The Kennicutt-Schmidt (KS) relation between the gas mass and star formation rate (SFR) describes the star formation regulation in disk galaxies. It is a function of gas metallicity, but the low metallicity regime of the KS diagram is poorly sampled. We have analyzed data for a representative set of extremely metal-poor galaxies (XMPs), as well as auxiliary data, and compared these to empirical and theoretical predictions. The majority of the XMPs possess high specific SFRs, similar to high redshift star-forming galaxies. On the KS plot, the XMP HI data occupy the same region as dwarfs, and extend the relation for low surface brightness galaxies. Considering the HI gas alone, a considerable fraction of the XMPs already fall off the KS law. Significant quantities of 'dark' H$_2$ mass (i.e., not traced by CO) would imply that XMPs possess low star formation efficiencies (SFE$_{\rm gas}$). Low SFE$_{\rm gas}$ in XMPs may be the result of the metal-poor nature of the HI gas. Alternatively, the HI reservoir may be largely inert, the star formation being dominated by cosmological accretion. Time lags between gas accretion and star formation may also reduce the apparent SFE$_{\rm gas}$, as may galaxy winds, which can expel most of the gas into the intergalactic medium. Hence, on global scales, XMPs could be HI-dominated, high specific SFR ($\gtrsim $ 10$^{-10}$~yr$^{-1}$), low SFE$_{\rm gas}$ ($\lesssim$ 10$^{-9}$~yr$^{-1}$) systems, in which the total HI mass is likely not a good predictor of the total H$_2$ mass nor of the SFR.

\end{abstract}


\keywords{galaxies: evolution -- galaxies: dwarfs}



\section{Introduction}

From observations of the Milky Way, Schmidt (1959) proposed that the star formation rate (SFR) and total gas volume density were related through a power law, reflecting the gas as the driver of star formation. The index was estimated to be approximately 2 from observations in the solar neighbourhood. More recently (Kennicutt 1989; Kennicutt 1998; Kennicutt \& Evans 2012), a relation was found to hold for global, disk-averaged total atomic HI plus molecular H$_2$ gas surface densities ($\Sigma _{\rm gas} \equiv \Sigma _{\rm H_2} + \Sigma _{\rm HI}$) and SFR surface densities ($\Sigma _{\rm SFR}$):

\begin{equation}
\Sigma _{\rm SFR} \propto \Sigma _{\rm gas}^n,
\end{equation}

\noindent where $n = 1.4 \pm 0.15$ in nearby spiral and starburst galaxies (hereinafter KS law). Subsequent observations of radial distributions and spatially resolved disks (individual star-forming regions) in nearby galaxies have demonstrated that the KS law is valid for $\Sigma _{\rm gas}$ down to a few M$_{\odot}$~pc$^{-2}$ (e.g., Bigiel et al. 2008, 2010; Daddi et al. 2010; Kennicutt \& Evans 2012; Wong et al. 2013). However, when the relation is investigated on kiloparsec-scales below this total gas surface density threshold, the data generally show a turnover (e.g., Bigiel et al. 2008; Elmegreen \& Hunter 2015). Below the threshold, the relation between the $\Sigma _{\rm SFR}$ and $\Sigma_{\rm gas}$ steepens, with an index $n \simeq $ 2 -- 3. By plotting $\Sigma_{\rm SFR}$ as a function of $\Sigma _{\rm H_2}$ alone, it is found that a correlation persists down to the lowest measurable H$_2$ column densities, resulting in an 'universal' molecular gas depletion timescale ($\tau_{\rm H_2} \equiv \Sigma_{\rm H_2}/\Sigma_{\rm SFR} \simeq$ 2.35~{\rm Gyr}; e.g., Bigiel et al. 2008, 2010). Notwithstanding, although $\Sigma_{\rm SFR}$ varies by approximately three to four orders of magnitude, $\Sigma_{\rm HI}$ varies by approximately one order of magnitude, showing a saturation limit at $\Sigma_{\rm HI} \simeq 9$~M$_{\odot}$ pc$^{-2}$, roughly at the position of the turnover seen in the data (e.g., Wong et al. 2013). It was proposed that this saturation limit signals a drop in the molecular gas fraction; below the threshold, the regime is HI-dominated. In this regime, although the HI dominates the H$_2$ molecules in column density, the H$_2$ gas depletion timescale remains $\tau_{\rm H_2}$ $\simeq$ 2~Gyr (e.g., Shruba et al. 2011). Above the threshold, the molecular gas is shielded from photodissociation, allowing an efficient phase transition from atomic to molecular gas. The saturation limit was demonstrated to depend on the metallicity (Krumholz, McKee \& Tumlinson 2008, 2009a, b). Recently, Elmegreen \& Hunter (2015) have modelled the relation between $\Sigma_{\rm SFR}$ and $\Sigma_{\rm HI}$ in the HI-dominated regime as a combination of three-dimensional gaseous gravitational processes and molecular hydrogen formation, with a 1\% efficiency per unit free-fall time. Elmegreen (2015) also describe a simple gas collapse model, at the local dynamical speed, where, in the HI-dominated regime, $\Sigma _{\rm SFR}$ $\propto$ $\Sigma _{\rm gas}^2$. The Krumholz (2013; hereinafter KMT+) analytical model computes the H$_2$ fraction and SFR as a function of the metallicity, gas surface density and the density of the stellar disk. Although this model assumes a, perhaps, unrealistic constant stellar density as the gas surface density varies, it reproduces the KS law above the threshold, the metallicity-dependent turnover and the steep correlation between $\Sigma_{\rm SFR}$ and $\Sigma_{\rm gas}$ in the HI-dominated regime, with a total gas depletion timescale ceiling of $\tau_{\rm gas} \equiv \Sigma_{\rm gas}/\Sigma_{\rm SFR} \simeq $ 100~Gyr.

This HI-dominated, low surface density region of the KS plot (i.e., $\Sigma _{\rm SFR}$ vs. $\Sigma _{\rm gas}$), below where the turnover occurs, is generally populated by HI-dominated outer regions of galaxies, low surface brightness (LSB) galaxies and low metallicity dwarf galaxies (e.g., Wyder et al. 2009; Fumagalli, Krumholz \& Hunt 2010; Bolatto et al. 2011; Shi et al. 2014; Cormier et al. 2014; Elmegreen \& Hunter 2015). At extremely low metallicities, molecule formation is affected (e.g., Glover \& Clark 2012b). CO is no longer an efficient tracer of the H$_2$ gas, due to possible dust depletion, a harder ionizing radiation field, and slower chemical reaction rates in a turbulent medium (e.g., Wolfire, Hollenbach \& McKee 2010; Shetty et al. 2011; Krumholz et al. 2008, 2009a, 2009b, 2013). Consequently, there is a large uncertainty associated with the CO to H$_2$ conversion factor, which has been suggested to depend on metallicity, and which can vary by over an order of magnitude (e.g., Leroy et al. 2011; Schruba et al. 2012; Bolatto, Wolfire \& Leroy 2013; Elmegreen et al. 2013; Cormier et al. 2014; Amor\'\i n et al. 2015; Shi et al. 2015; Hunt et al. 2015). Indeed, efforts to detect CO (as a proxy for the H$_2$ gas) in low metallicity dwarf galaxies have been largely unsuccessful; CO has only ever been detected in dwarfs with Z $ \gtrsim $ 0.1 Z$_{\odot}$ (e.g., Leroy et al. 2005; Bigiel et al. 2008; Schruba et al. 2012; Elmegreen et al. 2013; Cormier et al. 2014; Hunt et al. 2015; Shi et al. 2015). Dust, suggested as a better tracer of H$_2$ gas at low metallicities, also suffers from large uncertainties in the determination of the gas-to-dust ratio (e.g., R\'emy-Ruyer et al. 2013, 2014; Cormier et al. 2014; Shi et al. 2015; Hunt et al. 2015). Moreover, dust-based determinations of the H$_2$ mass in some low metallicity sources may differ by one to two orders of magnitude from that estimated from the CO emission (e.g., Cormier et al. 2014; Shi et al. 2014, 2015). In addition, observations of some of the lowest metallicity sources (e.g., Madden et al. 2012; R\'emy-Ruyer et al. 2013), and, in particular, recent observations of IZw 18 (Fisher et al. 2014; Hunt et al. 2014), suggest that dust may be depleted in some extremely low metallicity galaxies. In such low metallicity mediums, the [CII] line becomes a potential effective coolant of the gas (e.g., Wolfire, Hollenbach \& McKee 2010; Cormier et al. 2015). Observations of the [CII] line in some dwarf galaxies show line fluxes in large excess relative to the CO line, which has been interpreted as the presence of large amounts of 'dark' H$_2$ gas (e.g., Madden 2000; Leroy et al. 2005; Cormier et al. 2014, 2015). Indeed, such large molecular masses would imply that these galaxies are extremely inefficient at forming stars (e.g., Shi et al. 2014), or that a part of the gas is not involved in the star formation process (e.g., Cormier et a. 2014). It has also been suggested that, in extremely low metallicity galaxies, star formation will occur in collapsing cold atomic clouds rather than in molecular clouds, until the density reaches a high enough value that molecules begin to form in the cloud core, i.e., the star formation will proceed before the bulk of the atomic gas is converted into molecular gas (Glover \& Clark 2012a; Krumholz 2012).

The main aim of this paper is to adequately sample the low surface density region of the KS diagram, using information gathered for a local extremely metal-poor dwarf galaxy dataset (Morales-Luis et al. 2011; S\'anchez Almeida et al. 2016). This data, in combination with auxiliary literature data for dwarf galaxies and LSB galaxies, will allow us to test empirical and theoretical models of metallicity-dependent star formation, and empirical relations for molecular gas mass prediction.

The paper is organized as follows. Section~2 contains the detailed description of the extremely low metallicity dwarf galaxy and auxiliary data used in the analysis. This section contains the references and procedures used to compile the heterogeneous datasets used in the present study, and has been laid out to allow the reader to reproduce the results; it can be skipped without interrupting the thread of the argumentation in the other sections. Section~3 includes a qualitative comparative analysis of the data with respect to empirical and theoretical star formation predictions. Section~4 contains a discussion on empirical molecular mass scaling relations, and predicted molecular mass content for the extremely metal-poor dwarf galaxy dataset. Section~5 and 6, respectively, include a discussion of the results and conclusions. The Appendix contains supplementary information about some of the auxiliary and XMP data.

Throughout this paper, the cosmological parameters $\Omega _{\rm m}$ = 0.27, $\Omega _{\Lambda}$ = 0.73 and H$_{0}$ = 73 km s$^{-1}$ Mpc$^{-1}$ have been adopted.

\section{The Data}

\subsection{XMP Data}

Using the Sloan Digital Sky Survey (SDSS) Data Release 7 (DR7; Abazajian et al. 2009) and data from literature, Morales-Luis et al. (2011) assembled a comprehensive 140 source sample of local extremely metal-poor dwarf galaxies (XMPs). The galaxies were chosen so that they presented metallicities below one-tenth solar; explicitly, 12+log(O/H) $\lesssim$ 7.65 (see Morales-Luis et al. 2011 for details). Our present sample consists of 23 XMPs taken from this original XMP sample (hereinafter original XMP sample), which have both H${\alpha}$-derived SFR measurements and published HI data. In addition, the original Morales-Luis et al. (2011) XMP sample has been recently updated with 165 new XMPs (hereinafter new XMP sample), also from the SDSS DR7 (S\'anchez Almeida et al. 2016). The metallicity of the latter sample was estimated using the HII-CHI-mistry code by P\'erez-Montero (2014), which provides metallicities consistent with the direct method. From this updated XMP sample, 20 new XMPs have been included, which have both H${\alpha}$-derived SFR measurements and published HI data. 

Skillman et al. (2013) and James et al. (2015) postulated the existence of two types of extremely metal-deficient galaxies: 'starburst', whose star formation may be triggered by external causes, such as the infall of low metallicity gas, and 'quiescent', which may be metal-poor due to their small masses, and the existence of the mass-metallicity relationship (e.g., Lee et al. 2006). The XMPs in the present analysis correspond, mainly, to the Skillman et al. (2013) and James et al. (2015) 'starburst' category, i.e., they are bright for their metallicity (Filho et al. 2013). 




\setcounter{table}{0}

\begin{table*}

\tiny

\begin{center}

\begin{minipage}[c]{147mm}

\caption{Summary of the HI data for the new 20 XMPs in the local Universe, with H${\alpha}$-derived SFR measurements and published HI data.
}

\begin{tabular}{l c c c c c c c c c c }

\hline

Source & RA (J2000) & Dec (J2000) & $v_{\rm hel}$ & $v_{\rm peak}$ & $w_{50}$	  & $S_{\rm HI}$ &	log M$_{\rm HI}$ & 	$D$ & Ref \\			
       & $^h$ $^{m}$ $^{s}$ & $\degree$ $\arcmin$ $\arcsec$ & km s$^{-1}$   & km s$^{-1}$    & km s$^{-1}$ & Jy km s$^{-1}$ & M$_{\odot}$     & Mpc &      \\
 (1)   & (2)           & (3)            & (4)         & (5)            & (6)              & (7) & (8) & (9) & (10) \\

\hline
\hline

J093840.27+080809.8 & 09 38 40.29 & 08 08 9.86 & 3366 & \ldots	    & 103$\pm$10 & 1.70$\pm$0.06 & 8.94 		     & 46.7	& 1  \\
\ldots		        & 09 38 40.29 & 08 08 9.86 & 3366 & \ldots    & 103$\pm$10 & \ldots 		  & 8.60             & 46.7	& 2  \\
SBS0943+543			& 09 47 5.78 & 54 05 40.40 & 1594$\pm$20 & 3$\pm$0.8 &	 46	 & 0.21 		  & 7.55            & 26.7	& 3	 \\
UGC05347			& 09 57 16.52 & 04 31 37.22 & 2153 & \ldots    & 211$\pm$3	 & 6.24$\pm$0.09 & 9.15 		     & 30.8	& 1	 \\
J100642.44+511623.9	& 10 06 42.45 & 51 16 23.95 & 4901 & \ldots	    & 119		 & 0.34           &	8.59 		     & 69.9	& 4	 \\
J102344.95+270639.8 & 10 23 44.96 & 27 06 39.85 & 540  & \ldots    & 39$\pm$4   & 0.94$\pm$0.04 & 6.89             & 5.92	& 1	 \\
LSBC D640-13	    & 10 56 13.92 & 12 00 40.66 & 989  & \ldots    & 30$\pm$4	 & 1.84$\pm$0.04 & 7.89             & 13.3	& 1	 \\
\ldots		        & 10 56 13.92 & 12 00 40.66 & 990  & \ldots    & 42	         & 1.22 		  & 7.71  			& 13.3	& 5	 \\
SBS1119+586			& 11 22 37.77 & 58 19 42.68 & 1583$\pm$6 & \ldots & 52$\pm$13 & 1.32$\pm$0.36	& 8.34 			& 26.6	& 6  \\			
SBS1137+589 		& 11 40 32.09 & 58 38 32.03 & 1856$\pm$14 &	4$\pm$1.0 &	300	 & 0.73 		    & 8.26          & 32.5	& 3  \\
\ldots		        & 11 40 32.09 & 58 38 32.03 & 2032 & \ldots		& 110$\pm$17 & 0.85$\pm$0.39	& 8.33          & 32.5	& 6  \\
J121046.30+235454.9	& 12 10 46.31 & 23 54 54.94 & 2541 & \ldots	    & 129$\pm$9	 & 1.58$\pm$0.07	& 8.74          & 38.3	& 1	 \\
SBS1208+531 		& 12 11 0.68 & 52 49 56.91 & 973$\pm$6	& 11$\pm$3	& \ldots & 1.3	 		    & 7.94	        & 16.8	& 7	 \\
J121413.78+085429.7	& 12 14 13.80 & 08 54 29.73 & 1933 & \ldots	    & 95$\pm$3	 & 2.07$\pm$0.08	& 7.92          & 13.1	& 1 \\		
\ldots		        & 12 14 13.80 & 08 54 29.73 & 1933 & \ldots    & 95	$\pm$3	 & 1.89$\pm$0.07	& 7.88          & 13.1	& 8	\\
J122025.78+331431.7	& 12 20 25.79 & 33 14 31.75 & \ldots & 12	    & 33.9       & 0.72 		    & 7.82          & 19.8	& 9	 \\
J122712.80+073820.4	& 12 27 12.81 & 07 38 20.49 & 1170 & \ldots 	& 44$\pm$3	 & 1.28$\pm$0.05	& 7.71 		    & 13.1	& 1  \\
MCG+07-26-024		& 12 33 52.74 & 39 37 33.36 & \ldots &	113	    & 32.5       & 5.94             & 8.11          & 9.60	& 9	 \\
KUG1243+265			& 12 46 10.75 & 26 15 0.82 & 1878 & \ldots		& 54$\pm$6	 & 1.11$\pm$0.06	& 8.39          & 30.6	& 1	 \\
J140802.53+251507.2	& 14 08 2.53 & 25 15 7.30 & 3104 & \ldots	    & 23$\pm$11	 & 0.27$\pm$0.04	& 8.14          & 46.4	& 1	 \\
J141323.49+130443.9	& 14 13 23.50 & 13 04 43.99 & 4982 & \ldots 	& 113$\pm$6	 & 0.84$\pm$0.06	& 9.01          & 71.7	& 1	 \\	
J141710.66+104412.6	& 14 17 10.66 & 10 44 12.64 & 9737 & \ldots		& 238$\pm$11 & 0.94$\pm$0.10	& 9.61          & 136	& 1	 \\
J143238.20+121909.0	& 14 32 38.21 & 12 19 9.02 & 6703 & \ldots    & 164$\pm$44 & 0.70$\pm$0.08		& 9.17          & 95	& 1	 \\
J144038.18+035616.9	& 14 40 38.19 & 03 56 16.98 & 8549 & \ldots    & 128$\pm$32 & 0.27$\pm$0.06	& 8.95          & 118.6	& 1	 \\

\hline


\end{tabular}
\\
Columns: \\
Col. 1: Source name. \\
Col. 2: Right ascension (J2000). \\
Col. 3: Declination (J2000). \\
Col. 4: Heliocentric velocity (and error). \\
Col. 5: HI velocity profile peak (and error). \\
Col. 6: HI line width at 50\% of the peak flux density (and error). \\
Col. 7: HI integrated flux density (and error). \\
Col. 8: Logarithm of the derived (total) HI mass. \\
Col. 9: Virgocentric infall-corrected Hubble flow distance from the NED. \\
Col. 10: Reference for the HI data. \\
\\
Code in column 10, and instrument/survey: 
\\
1 -- Haynes et al. (2011) -- Arecibo Legacy Fast ALFA Survey (ALFALFA) \\
2 -- Stierwalt et al. (2009) -- ALFALFA \\
3 -- Huchtmeier, Gopal \& Petrosian (2005) -- Effelsberg Radio Telescope \\
4 -- Kreckel et al. (2011) -- Westerbork Synthesis Radio Telescope (WSRT) \\
5 -- Eder \& Schombert (2000) -- Arecibo Radio Telescope \\
6 -- Thuan et al. (1999) -- Very Large Array (VLA) \\
7 -- Huchtmeier et al. (2007) -- Effelsberg Radio Telescope \\
8 -- Kent et al. (2008) -- ALFALFA \\
9 -- Kova\v {c}, Oosterloo \& van der Hulst (2009) -- WSRT \\

\end{minipage}

\end{center}

\end{table*}








The SFRs that are used in the present analysis were obtained from the Max Planck-John Hopkins University (MPA-JHU) DR7 release of spectrum measurements\footnote{http://www.mpa-garching.mpg.de/SDSS/DR7/}. SFRs are inferred from extinction-corrected H${\alpha}$ emission-line luminosities and extrapolated to the whole galaxy, according to the procedure outlined in Brinchmann et al. (2004). The MPA-JHU SFR measurements are in good agreement with those resulting from dust-corrected GALEX\footnote{Galaxy Evolution Explorer} measurements (Salim et al. 2007). 

The XMP optical radii ($r_{\rm opt}$) used for the surface density estimations were obtained from the SDSS $r$-band images, or, when not available, the Digitized Sky Survey (DSS)-II $R$-band images, measured up to the 25 mag arcsec$^{-2}$ isophote. Because XMPs generally show very diffuse, irregular optical morphology, the optical radii have not been corrected for projection effects. The optical radii were found to vary by approximately an order of magnitude across the XMP sample: $r_{\rm opt} \simeq $ 0.2 -- 5~kpc. Distances were corrected for Virgocentric infall, and were obtained from the NASA/IPAC Extragalactic Database\footnote{http:/ned.ipac.caltech.edu/} (NED). 

Filho et al. (2013) presented new single-dish Effelsberg HI measurements, and low resolution/single-dish and/or interferometric HI data from literature for the original XMP sample. The HI data for the new XMP sample are contained in Table~1. The total HI masses for the new XMP sample were calculated utilizing the same procedure as in Filho et al. (2013). When multiple HI entries are present, the most appropriate total HI measurement was chosen for the subsequent data analysis. It is found that the total HI masses of the XMPs are at least an order of magnitude larger than the stellar masses, while the HI radii are at least three times the optical radii (see Filho et al. 2013 and references therein). HI masses and HI surface densities, within the optical radius, have been derived in a very simplified manner, as follows. Interferometric studies of star-forming dwarf galaxies have found that the HI gas generally follows an exponential profile, $\Sigma_{\rm HI}$($x$) $\propto$ exp ($-x/h$), where $h$ is the scale-length and $x$ is the galactocentric radius (e.g., van Zee, Skillman \& Salzer 1998; van Zee, Salzer \& Skillman 2001; Lelli, Verheijan \& Fraternali 2014a, 2014b; Roychowdhury et al. 2011, 2014; Elmegreen \& Hunter 2015). In addition, it has been demonstrated that the HI gas is more centrally concentrated in blue compact dwarf galaxies (BCDs; which are also more compact in the optical: $r_{\rm opt} < $ 3~kpc) than in dwarf irregular galaxies (dIrrs; which are also more extended in the optical: $r_{\rm opt} \gtrsim $ 3~kpc). According to this size criterium, only four of the sample sources are XMP 'dIrrs'. Assuming that the HI-to-optical radius is constant for each 'BCD'/'dIrr' subgroup, $h$ = 3/2 $r_{\rm opt}$ for XMP 'BCDs' and $h$ = 2 $r_{\rm opt}$ for XMP 'dIrrs' have been adopted (Fig.~11 in van Zee, Skillman \& Salzer 1998, and Fig.~10 in van Zee, Salzer \& Skillman 2001). HI gas surface densities and HI masses, within the optical radius, have been derived from the total HI mass assuming the exponential profile for the HI gas, and the scale-length values as defined above. H$_2$ masses, within the optical radius, have been derived from the metallicity, SFR and/or HI mass, using the empirical mass scaling relations given in Section~4, and subsequently used to compute the H$_2$ surface gas densities, within the optical radius. Total (H$_2$+HI) gas surface densities, within the optical radius, have also been derived. These values do not include helium; including helium would increase $\Sigma _{\rm gas}$ by $\simeq$ 0.134 in log, and would not significantly alter the results (Sect.~5). The errors in $\Sigma_{\rm HI}$ and $\Sigma_{\rm SFR}$ were estimated considering the errors in HI mass (see Filho et al. 2013 for details), a 20\% error in the determination of the galaxy optical radius, and a 1$\sigma $ error in the SFR from the MPA-JHU dataset. When a quoted error in the HI integrated flux density ($S_{\rm HI}$) is not available (Table~1), a 5\% error has been assumed, comparable to the errors quoted for the other XMPs.

Table~2 provides the relevant auxiliary, and SFR, HI and total gas surface density data for the sample of 43 XMPs, 23 from the original sample and 20 from the new sample.

Several XMPs with other literature data have also been included in the analysis. Sextans A and ESO 146-G14 are XMPs belonging to the original XMP sample. However, because MPA-JHU SFRs estimates are not available, their integrated HI gas and SFR surface densities have been derived as follows: an integrated SFR surface density value from Dolm-Palmer et al. (2002; their Fig.~8; $\Sigma _{\rm SFR} \simeq $ 0.015 M$_{\odot}$~yr$^{-1}$~kpc$^{-2}$) has been adopted for Sextans A, and an average (integrated over the galaxy and over the galaxy lifetime) SFR value from Bergvall \& R\"onnback (1995; SFR $\simeq$  0.3 M$_{\odot}$~yr$^{-1}$) has been adopted for ESO 146-G14. Atomic gas masses are from Filho et al. (2013), and optical source sizes are from the present work. Also included are resolved measurements for individual star-forming regions in Sextans A and ESO 146-G14 (Shi et al. 2014), where $\Sigma _{\rm H_2}$ (Spitzer and Herschel data), $\Sigma _{\rm HI}$ (VLA\footnote{Very Large Array} and ATCA\footnote{Australia Telescope Compact Array} data) and $\Sigma _{\rm SFR}$ (Herschel, Spitzer, and GALEX data) are sampled within $r_{\rm dust} \simeq $ 0.2 -- 1~kpc (the dust aperture).

The Dwarf Galaxy Survey (DGS; Madden et al. 2013) contains 18 sources in common with the original XMP sample, only three of which possess CO observations that yield H$_2$ mass upper limits (Cormier et al. 2014): SBS 0335-052, IZw 18 and VII Zw 403 (UGC 6456). The integrated measurements for these XMPs are included in the subsequent analysis, where $\Sigma _{\rm H_2}$ (Spitzer, APEX\footnote{Atacama Pathfinder Experiment}, Mopra 22-m and IRAM\footnote{Institut de Radioastronomie Millim\'etrique} 30-m data), $\Sigma _{\rm HI}$ (see Madden et al. 2013 for details) and $\Sigma _{\rm SFR}$ (Spitzer data) are sampled within $r_{\rm dust} \simeq $ 50 -- 90\arcsec~or $r_{\rm CO} \simeq $ 10 -- 30\arcsec~(the dust and CO aperture, respectively).




\setcounter{table}{1}


\begin{table*}

\tiny


\begin{center}

\begin{minipage}[c]{162mm}

\caption{Summary of the auxiliary, and SFR, HI and total gas surface density data for the 43 XMPs in the local Universe, with H${\alpha}$-derived SFR measurements and published HI data.
}

\begin{tabular}{l c c c c c c c c c c c c}

\hline

Source & $r_{\rm opt}$ & log SFR   & log $\Sigma _{\rm SFR}$	  & log $\Sigma_{\rm HI}$ & log M$_{\rm H_2}^{\rm Shi}$ & log $\Sigma_{\rm gas}^{\rm Shi}$ & log M$_{\rm H_2}^{\rm Amorin}$ & log $\Sigma_{\rm gas}^{\rm Amorin}$ & log M$_{\rm H_2}^{\rm universal}$ & log $\Sigma_{\rm gas}^{\rm universal}$ & log M$_{\rm H_2}^{\rm short}$ & log $\Sigma_{\rm gas}^{\rm short}$\\			
       & kpc           &  M$_{\odot}$ yr$^{-1}$ & M$_{\odot}$ yr$^{-1}$ kpc$^{-2}$ & M$_{\odot}$ pc$^{-2}$ & M$_{\odot}$ & M$_{\odot}$ pc$^{-2}$ & M$_{\odot}$ & M$_{\odot}$ pc$^{-2}$ & M$_{\odot}$ & M$_{\odot}$ pc$^{-2}$ & M$_{\odot}$ & M$_{\odot}$ pc$^{-2}$ \\
 (1)   & (2)           & (3)            & (4)         & (5)            & (6)              & (7) & (8) & (9)  & (10)	& (11) & (12) & (13) \\

\hline
\hline

Original XMP Sample & & & & & & & & & & & & \\

\hline
 
J0119-0935   &    1.31   &   -2.14   &   -2.61   &    1.03  & 8.54   &    1.88   &    9.20   &    2.49    &   7.49    &   1.22   &    7.13    &   1.12  \\
HS0122+0743  &    6.54   &   -1.17   &   -3.01   &    0.68  & 9.59   &    1.53   &    10.02  &    1.92    &   8.48    &   0.85   &    8.11    &   0.76  \\
J0126-0038   &    1.86   &   -2.93   &   -3.71   &    1.22  & 9.03   &    2.06   &    9.49   &    2.48    &   6.68    &   1.23   &    6.32    &   1.22  \\
J0133+1342   &    0.88   &   -1.39   &   -1.49   &    0.73  & 7.89   &    1.58   &    8.05   &    1.71    &   8.25    &   1.90   &    7.89    &   1.58  \\
J0204-1009   &    1.93   &   -1.54   &   -2.33   &    1.73  & 9.57   &    2.57   &    10.36  &    3.31    &   8.10    &   1.81   &    7.74    &   1.76 \\
J0301-0052   &    0.67   &   -1.99   &   -1.87   &    2.10  & 9.02   &    2.94   &    9.46   &    3.34    &   7.64    &   2.19   &    7.28    &   2.14 	\\
HS0822+03542 &    0.39   &   -1.99   &   -1.41   &    0.88  & 7.34   &    1.72   &    7.70   &    2.05    &   7.64    &   1.99   &    7.28    &   1.67 \\
IZw18      	 &    0.62   &   -1.74   &   -1.55   &    1.66  & 8.52   &    2.50   &    9.39   &    3.31    &   7.90    &   2.04   &    7.53    &   1.87 \\ 
J0940+2935   &    0.76   &   -2.09   &   -2.09   &    0.76  & 7.80   &    1.61   &    7.80   &    1.61    &   7.54    &   1.39   &    7.18    &   1.15 \\
KUG1013+381  &    1.38   &   -1.46   &   -1.96   &    1.00  & 8.55   &    1.85   &    8.81   &    2.07    &   8.18    &   1.55   &    7.82    &   1.32 \\
UGCA211      &    1.89   &   -2.20   &   -2.99   &    0.80  & 8.63   &    1.65   &    8.94   &    1.92    &   7.43    &   0.94   &    7.06    &   0.87 \\
HS1033+4757  &    0.89   &   -1.89   &   -2.02   &    1.54  & 8.71   &    2.38   &    8.90   &    2.54    &   7.74    &   1.75   &    7.38    &   1.64  \\
J1105+6022   &    1.99   &   -1.53   &   -2.35   &    1.03  & 8.90   &    1.87   &    9.14   &    2.08    &   8.11    &   1.32   &    7.75    &   1.18  \\
J1201+0211   &    0.45   &   -1.99   &   -1.52   &    1.05  & 7.62   &    1.89   &    7.83   &    2.08    &   7.64    &   1.91   &    7.28    &   1.62 \\
SBS1211+540  &    0.68   &   -1.94   &   -1.83   &    1.11  & 8.04   &    1.95   &    8.11 	 &    2.01    &   7.69    &   1.67   &    7.33    &   1.44  \\
J1215+5223   &    0.43   &   -2.82   &   -2.33   &    0.96  & 7.49   &    1.80   &    7.76   &    2.04    &   6.80    &   1.30   &    6.43    &   1.14  \\
VCC0428      &    0.49   &   -1.92   &   -1.53   &    1.17  & 7.82   &    2.01   &    7.84   &    2.03    &   7.71    &   1.92   &    7.35    &   1.65  \\
J1230+1202   &    0.55   &   -1.69   &   -1.40   &    1.26  & 8.02   &    2.11   &    7.95   &    2.04    &   7.95    &   2.04   &    7.58    &   1.76 \\
GR8          &    0.27   &   -3.16   &   -2.27   &    0.90  & 7.04   &    1.75   &    6.98   &    1.62    &   6.45    &   1.31   &    6.09    &   1.13 \\ 
HS1442+4250  &    2.36   &   -2.08   &   -3.05   &    0.80  & 8.82   &    1.65   &    9.19   &    1.98    &   7.55    &   0.92   &    7.19    &   0.86 \\
J2053+0039   &    1.89   &   -1.73   &   -2.51   &    1.65  & 9.48   &    2.50   &    10.30  &    3.26    &   7.91    &   1.72   &    7.54    &   1.68  \\ 
J2104-0035   &    0.64   &   -2.01   &   -1.86   &    1.67  & 8.56   &    2.52   &    9.62   &    3.51    &   7.62    &   1.90   &    7.26    &   1.78 \\
PHL293B      &    0.74   &   -1.52   &   -1.48   &    1.32  & 8.33   &    2.16   &    8.49   &    2.30    &   8.12    &   1.99   &    7.76    &   1.73 \\ 

\hline

\end{tabular}

\end{minipage}

\end{center}

\end{table*}














\setcounter{table}{1}


\begin{table*}

\tiny


\begin{center}

\begin{minipage}[c]{162mm}

\caption{Summary of the auxiliary, and SFR, HI and total gas surface density data for the 43 XMPs in the local Universe, with H${\alpha}$-derived SFR measurements and published HI data (continued).
}

\begin{tabular}{l c c c c c c c c c c c c}

\hline

Source & $r_{\rm opt}$ & log SFR   & log $\Sigma _{\rm SFR}$	  & log $\Sigma_{\rm HI}$ & log M$_{\rm H_2}^{\rm Shi}$ & log $\Sigma_{\rm gas}^{\rm Shi}$ & log M$_{\rm H_2}^{\rm Amorin}$ & log $\Sigma_{\rm gas}^{\rm Amorin}$ & log M$_{\rm H_2}^{\rm universal}$ & log $\Sigma_{\rm gas}^{\rm universal}$ & log M$_{\rm H_2}^{\rm short}$ & log $\Sigma_{\rm gas}^{\rm short}$\\			
       & kpc           &  M$_{\odot}$ yr$^{-1}$ & M$_{\odot}$ yr$^{-1}$ kpc$^{-2}$ & M$_{\odot}$ pc$^{-2}$ & M$_{\odot}$ & M$_{\odot}$ pc$^{-2}$ & M$_{\odot}$ & M$_{\odot}$ pc$^{-2}$ & M$_{\odot}$ & M$_{\odot}$ pc$^{-2}$ & M$_{\odot}$ & M$_{\odot}$ pc$^{-2}$ \\
 (1)   & (2)           & (3)            & (4)         & (5)            & (6)              & (7) & (8) & (9)  & (10)	& (11) & (12) & (13) \\

\hline
\hline

New XMP Sample & & & & & & & & & & & & \\

\hline

J093840.27+080809.8  & 1.30   &   -1.13   &   -1.85   &    1.84   &    9.34   &    2.68   &    10.15   &    3.43   &    8.23   &    2.00    &   7.87   &    1.92  \\
SBS0943+543		&	0.81    &  -1.59   &   -1.91   &    0.86  &     7.95   &    1.71   &    8.21   &    1.94   &    7.77   &    1.55    &   7.41   &    1.29  \\  
UGC05347		&	 5.61   &   -0.93  &    -2.92   &    0.63  &     9.41   &    1.48  &     9.93   &    1.95  &     8.44  &     0.85   &    8.07  &     0.74  \\ 
J100642.44+511623.9	& 1.86  &    -0.75  &    -1.79   &    1.18  &     8.99  &     2.02  &     9.38  &     2.37 &      8.61  &     1.72  &     8.25 &      1.50 \\  
J102344.95+270639.8 & 0.34   &   -2.60  &    -2.15   &    0.96   &    7.29   &    1.81   &    7.60    &   2.08   &    6.76   &    1.40   &    6.40   &    1.21  \\  
LSBC D640-13	 &   0.90  &    -1.93  &    -2.34   &    1.11   &    8.29    &   1.95    &   8.38   &    2.03    &   7.43   &    1.37   &    7.07   &    1.24  \\ 
SBS1119+586		&	1.29    &  -1.43  &    -2.14   &    1.25    &   8.74    &   2.09     &  8.96   &    2.29   &    7.94   &    1.53   &    7.57    &   1.39 \\   		
SBS1137+589 	&	0.77    &  -2.68   &   -2.95  &     1.61   &    8.66    &   2.46   &    8.94    &   2.71   &    6.69  &     1.64   &    6.32   &    1.62 \\ 
J121046.30+235454.9	& 1.76   &   -1.66   &   -2.65   &    1.37   &    9.14    &   2.22   &    9.61    &   2.64   &    7.70    &   1.46    &   7.34  &     1.41 \\  
SBS1208+531		  &   0.49   &   -2.20   &   -2.08   &    1.69   &    8.34    &   2.53    &   8.39   &    2.58   &    7.16    &   1.83    &   6.80   &    1.76 \\  
J121413.78+085429.7	&  0.65    &  -1.74   &   -1.87   &    1.42    &   8.32    &   2.27    &   8.76    &   2.66   &    7.62   &    1.76    &   7.26   &    1.60 \\ 	
J122025.78+331431.7 & 0.60   &   -2.12   &   -2.17   &    1.39    &   8.22   &    2.24   &    9.04    &   3.00    &   7.24   &    1.60   &    6.88   &    1.50 \\  
J122712.80+073820.4	& 0.57   &   -3.24    &  -3.25   &    1.32    &   8.11   &    2.17   &    8.34   &    2.37    &   6.12    &   1.35   &    5.76   &    1.33 \\  
MCG+07-26-024	&	  1.12   &   -2.14   &   -2.73   &    1.14   &    8.51   &    1.99   &    8.79    &   2.24    &   7.23    &   1.26    &   6.86   &    1.20 \\   
KUG1243+265		&	 1.67   &   -1.18   &   -2.12   &    1.07   &    8.79   &    1.92    &   8.95   &   2.05   &    8.18    &   1.46    &   7.82    &   1.29 \\   
J140802.53+251507.2 & 1.18   &   -1.59   &   -2.23   &    1.12   &    8.54   &    1.97   &   9.05   &    2.43   &    7.78   &    1.43   &    7.41    &   1.28 \\   
J141323.49+130443.9 &	2.00  &    -0.94   &   -2.04    &   1.54    &   9.41   &    2.38  &     9.69   &    2.63   &    8.42  &     1.74   &    8.06  &     1.64 \\  	
J141710.66+104412.6	& 3.30 &     -0.73   &   -2.26   &    1.56   &    9.87   &    2.40   &    10.40   &   2.89   &    8.63  &     1.69  &     8.27   &    1.62 \\  
J143238.20+121909.0	& 2.76    &  -0.73  &    -2.11  &     1.41  &     9.57   &    2.26   &    10.09   &   2.74   &    8.63  &     1.64  &     8.27   &    1.53 \\  
J144038.18+035616.9	&  3.74   &   -1.49  &    -3.14   &    0.79   &    9.21    &   1.63   &    9.61   &    2.00   &    7.87 &      0.89  &     7.51   &    0.84 \\   

\hline

\end{tabular}


Columns: \\
Col. 1: Source name. \\
Col. 2: Optical radius from the SDSS $r$-band image, or, when not available, the DSS-II $R$-band image, and measured at 25 mag arcsec$^{-2}$. \\
Col. 3: Median estimate of the logarithm of the total SFR from the MPA-JHU DR7 release of spectrum measurements, obtained from the extinction-corrected H${\alpha}$ emission-line luminosities and integrated over the whole galaxy, according to the procedure outlined in Brinchmann et al. (2004). \\
Col. 4: Logarithm of the SFR surface density within the optical radius. \\
Col. 5: Logarithm of the HI gas surface density within the optical radius. \\
Col. 6: Derived logarithm of the H$_2$ mass, within the optical radius, assuming the scaling relation of Shi et al. (2014; Sect.~4). \\
Col. 7: Derived logarithm of the total (H$_2$+HI) gas surface density, within the optical radius, assuming the scaling relation of Shi et al. (2014; Sect.~4). \\
Col. 8: Derived logarithm of the H$_2$ mass, within the optical radius, assuming the scaling relation of Amor\'\i n et al. (2015; Sect.~4). \\
Col. 9:  Derived logarithm of the total (H$_2$+HI) gas surface density, within the optical radius, assuming the scaling relation of Amor\'\i n et al. (2015; Sect.~4). \\ 
Col. 10: Derived logarithm of the H$_2$ mass, within the optical radius, assuming the 'universal' H$_2$ depletion timescale of 2.35~Gyr (Bigiel et al. 2008, 2010; Sect.~4). \\
Col. 11: Derived logarithm of the total (H$_2$+HI) gas surface density, within the optical radius, assuming the 'universal' H$_2$ depletion timescale of 2.35~Gyr (Bigiel et al. 2008, 2010; Sect.~4). \\ 
Col. 12: Derived logarithm of the H$_2$ mass, within the optical radius, assuming the 'short' H$_2$ depletion timescale of 1~Gyr (Sect.~4). \\
Col. 13: Derived logarithm of the total (H$_2$+HI) gas surface density, within the optical radius, assuming the 'short' H$_2$ depletion timescale of 1~Gyr (Sect.~4). \\ 

\end{minipage}

\end{center}


\end{table*}









We caution that the use of standard calibrations to infer the SFR from the integrated light of a stellar population, generally applied at higher SFRs, may lead to systematic errors when applied at low SFRs (SFR $\lesssim$ 10$^{-3}$ M$_{\odot}$ yr$^{-1}$), as is the case in some of the XMP data. The use of the H$\alpha$ emission to infer the SFR can be particularly problematic, due to the stochastic nature of the star formation, and the short time-scales over which the integration is performed (da Silva, Fumagalli \& Krumholz 2014). Note that this may also affect some of the auxiliary data (Sect.~2.2).

In the following figures (Figure~1 -- 5), it is apparent that there are a few XMPs with low specific SFRs (sSFR $\equiv $ SFR/M$_{\ast}$; sSFR $\lesssim$ 10$^{-10}$~yr$^{-1}$), although they appear to be a continuation of other general XMP properties. These XMPs also generally possess the lowest SFRs (see above) and SFR surface densities. It is to be noted that, like the higher sSFR XMPs, these XMPs are 'starburst' (Skillman et al. 2013; James et al. 2015), according to the metallicity-luminosity relation (Filho et al. 2013). Closer inspection of the SDSS and MPA-JHU data reveals that these low sSFR XMPs are generally multi-knot star-forming sources, where the automated estimation of the (MPA-JHU) integrated SFR has likely been unsuccessful; the real integrated SFR is presumably higher. These (few) low sSFR XMPs are included in the following plots for completeness, but no conclusions are drawn regarding their SFR properties.

\subsection{Auxiliary Data}

In the subsequent analysis, the XMP data have been compared with the resolved measurements for individual star formation regions in (1) the low-metallicity (Z $\sim$ 0.2 Z$_{\odot}$) dIrr Small Magellanic Cloud (SMC; Bolatto et al. 2011), where $\Sigma _{\rm H_2}$ (Spitzer data), $\Sigma_{\rm HI}$ (ATCA and Parkes 64-m data) and $\Sigma _{\rm SFR}$ (H${\alpha }$ data) are sampled within $r_{\rm opt} \simeq $ 1~kpc and (2) the Northwestern region (cloud A) of the metal-poor (12+log(O/H) $\sim$ 7.80) dIrr Wolf-Lundkmark-Melotte (WLM; Elmegreen et al. 2013), where $\Sigma _{\rm H_2}$ (Spitzer data), $\Sigma _{\rm HI}$ (VLA data) and $\Sigma _{\rm SFR}$ (H${\alpha}$ and GALEX data) are sampled within $r_{\rm CO} \simeq $ 50~pc. Included also is a comparison with radial distribution values for (3) LSB galaxies (Wyder et al. 2009), where $\Sigma _{\rm HI}$ (VLA and WSRT\footnote{Westerbork Synthesis Radio Telescope} data) and $\Sigma _{\rm SFR}$ (GALEX data) are sampled within $r_{\rm UV} \simeq $ 20 -- 130\arcsec~(the ultraviolet aperture). Further included is a comparison with integrated values for (4) the Milky Way (Kennicutt \& Evans 2012), where $\Sigma _{\rm H_2}$ and $\Sigma _{\rm HI}$ (see Kalberla \& Kerp 2009 for details), and $\Sigma _{\rm SFR}$ (see Chomiuk \& Povich 2011 for details) are sampled within $r_{\rm opt} \simeq $ 13.5~kpc, (5) the metal-poor dIrr WLM, where the total HI mass (log(M$_{\rm HI}$/M$_{\odot}$) $\simeq $ 7.8) and distance ($D \simeq $ 0.95~Mpc) are from Kepley et al. (2007), the source size from the NED, and the integrated H${\alpha}$-derived SFR (SFR $\simeq $ 0.002 M$_{\odot}$~yr$^{-1}$) from Hunter, Elmegreen \& Ludka (2010), and (6) several dwarf galaxies from the DGS (Cormier et al. 2014) with a range in metallicities (7.7 $\lesssim$ 12+log(O/H) $\lesssim$ 8.4), where $\Sigma _{\rm H_2}$ (Spitzer, APEX, Mopra 22-m and IRAM 30-m data), $\Sigma _{\rm HI}$ (see Madden et al. 2013 for details) and $\Sigma _{\rm SFR}$ (Spitzer data) are sampled within $r_{\rm dust} \simeq $ 50 -- 90\arcsec~or $r_{\rm CO} \simeq $ 10 -- 30\arcsec. 


Appropriate dust-to-gas and CO-to-gas conversion factors have been applied to the data (see individual references for details). The SFRs for the LSB galaxies (Wyder et al. 2009) have been recalibrated to the values of Kennicutt \& Evans (2012).

\section{KS Plot}

Figure~1a contains the SFR surface density as a function of the total (solid symbols) or HI (open symbols) gas surface density for the auxiliary data (Sect.~2.2), plus the resolved observations for the XMPs Sextans A (red open and solid squares; Shi et al. 2014) and ESO 146-G14 (red open and solid circles; Shi et al. 2014), and the integrated measurements for the XMPs also contained in the DGS: SBS 0335-052, IZw 18 and VII Zw 403 (encircled brown open and solid triangles; Cormier et al. 2014). Figure~1b contains the SFR surface density as a function of the HI gas surface density, integrated over the optical radii, for the 43 sample XMP galaxies (blue open circles; Sect~2.1), plus the integrated values for the XMPs Sextans A (red open square; Sect.~2.1) and ESO 146-G14 (red open circle; Sect.~2.1), and for the metal-poor dIrr WLM (dark green open triangle; Sect.~2.2). In both figures, the KS law is plotted (magenta solid line; Kennicutt \& Evans 2012), as well as the observed HI relation for the Faint Irregular Galaxies GMRT Survey (FIGGS; yellow solid line; Roychowdhury et al. 2014), and the observed (H$_2$+HI) relation for starburst galaxies (purple solid line; Daddi et al. 2010). 

Figure~1a and b also include an analytical KMT+ model (black solid line; Krumholz 2013), corresponding to a model with a clumping factor of $f_c$ = 5, a volume density of stars and dark matter within the gas disk of $\rho _{\rm sd}$ = 0.1, and a metallicity of Z = 0.1 Z$_{\odot}$, which is representative of the conditions expected in XMP galaxies. Lines of constant (total or HI) gas depletion timescale, $\tau \equiv \tau _{\rm gas} \equiv \tau _{\rm H_2+HI}$ or $\tau \equiv \tau _{\rm HI}$, respectively, are also included (grey solid and dotted lines).



\setcounter{figure}{0}

\begin{figure}
\begin{center}

\includegraphics[width=7.5cm]{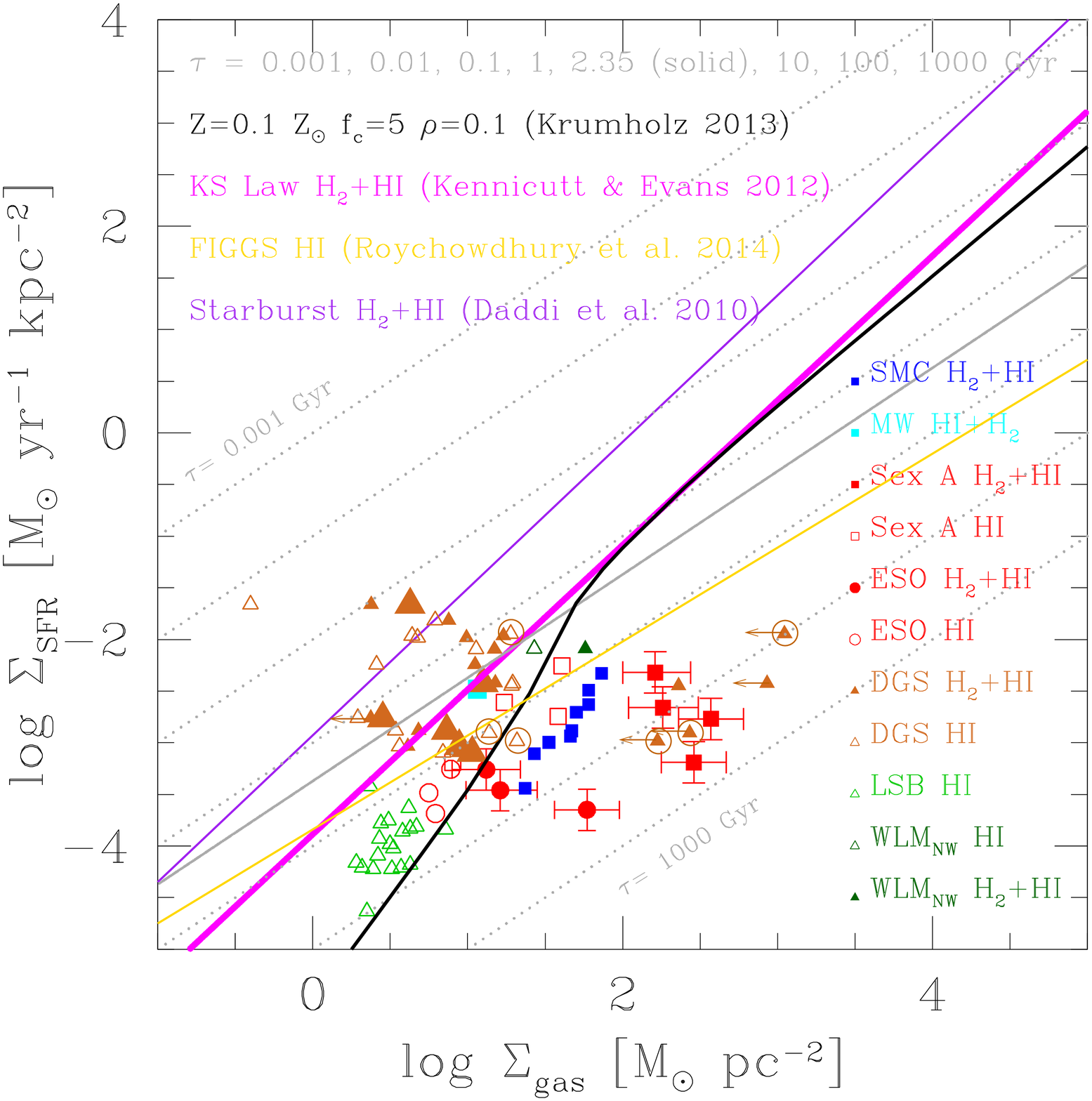}
(a)
\includegraphics[width=7.5cm]{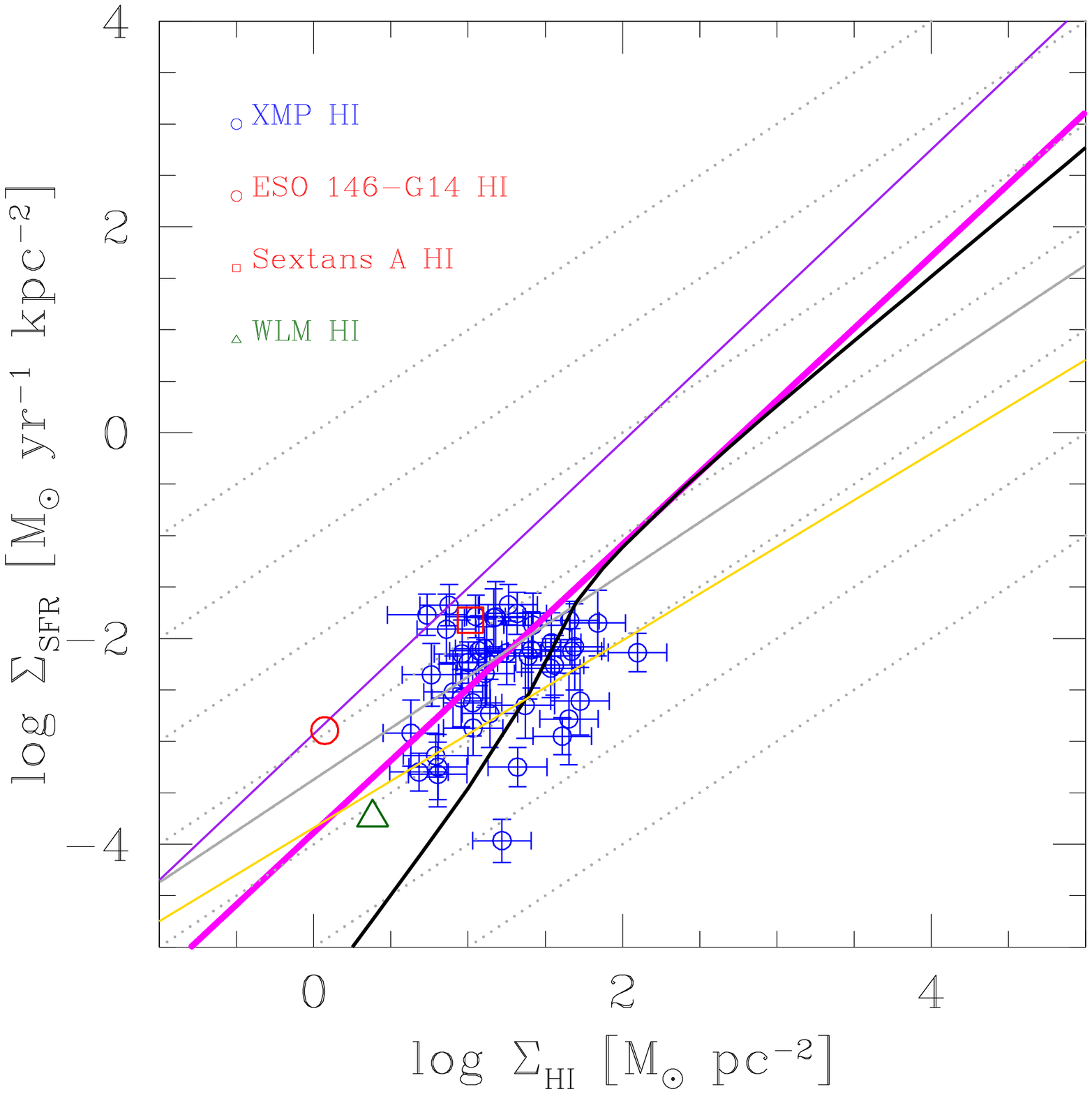}
(b)
 
\caption{(a) The SFR surface density as a function of the total (solid symbols) or HI (open symbols) gas surface density for the auxiliary data, plus resolved observations for the XMPs Sextans A (red open and solid squares; Shi et al. 2014) and ESO 146-G14 (red open and solid circles; Shi et al. 2014), and the integrated measurements for the XMPs also contained in the DGS: SBS 0335-052, IZw 18 and VII Zw 403 (encircled brown open and solid triangles; Cormier et al. 2014). The auxiliary data include spatially resolved observations of the SMC (blue solid squares; Bolatto et al. 2011) and WLM (dark green open and solid triangles; Elmegreen et al. 2013), integrated values for the Milk Way (cyan solid square; Kennicutt \& Evans 2012) and several DGS dwarf galaxies (brown open and solid triangles, where the smaller triangles correspond to CO-, and the larger triangles to dust-, derived H$_2$ estimations; Cormier et al. 2014), and radial distribution values for LSB galaxies (light green open triangles; Wyder et al. 2009). (b) The SFR surface density as a function of the HI gas surface density, integrated over the optical radii, for the 43 sample XMPs (blue open circles, Sect.~2.1), plus the integrated values for the XMPs Sextans A (red open square; Sect.~2.1) and ESO 146-G14 (red open circle; Sect.~2.1), and for the metal-poor dIrr WLM (dark green open triangle; Sect.~2.2). The magenta solid line corresponds to the KS law (Kennicutt \& Evans 2012), in yellow the observed HI relation for FIGGS galaxies (Roychowdhury et al. 2014), in purple the observed (H$_2$+HI) relation for starburst galaxies (Daddi et al. 2010), and in black a representative KMT+ model (Krumholz 2013). Grey lines of constant (total or HI) gas depletion timescale are also included: (from top to bottom) $\tau $ = 0.001, 0.01, 0.1, 1, 2.35 (solid), 10, 100 and 1000 Gyr. See the caveat on low SFRs and low sSFRs in Section~2.}

\end{center}
\end{figure}


\subsection{DGS Data}

The DGS data show a large dispersion on the KS plot; the molecular gas content (brown solid triangles; Fig.~1a) is anywhere from half a magnitude less, to an order of magnitude larger than the atomic gas content (brown open triangles; Fig.~1a). There are a few dwarfs which exhibit short total gas depletion timescales ($\tau _{\rm gas} \lesssim$ 1~Gyr), that are located in the region of the KS plot typically dominated by merging/starburst systems (purple solid line; Fig. 1a). Indeed two of these sources (Haro 11 and NGC 5253) show indication of having undergone an accretion event (see Cormier et al. 2014 and references therein for details). These dwarfs may represent sources where the merger is in a more advanced stage, with the gas being quickly depleted. At the other extreme, there are several sources which may possess larger quantities of total gas (only H$_2$ upper limits are available), but exhibit low star formation efficiencies (SFE$_{\rm gas} \equiv \Sigma _{\rm SFR}$/$\Sigma _{\rm gas}$), up to 10 -- 100 times lower than the KS law predictions. Two of these sources, the XMPs SBS 0335-052 and IZw 18 (see Filho et al. 2015 and references therein for details; see also the Appendix), are documented interacting galaxies, and UM 461 is suspected to be interacting (Doublier et al. 1999). These sources could represent a transient, perhaps less advanced phase of the interaction process, whereby the HI (and possibly the H$_2$) content is high, but they have not begun forming stars at the appropriate rate.



\subsection{XMP Data}

For the XMPs, the HI star formation efficiency:

\begin{equation}
{\rm SFE_{HI}} \equiv \Sigma _{\rm SFR}/\Sigma _{\rm HI},
\end{equation}

\noindent or equivalently, its inverse, the HI gas depletion timescale:

\begin{equation}
\tau _{\rm HI} \equiv \Sigma _{\rm HI}/\Sigma _{\rm SFR},
\end{equation}

\noindent shows a broad range of values: 10$^{-11}$ -- 10$^{-8}$~yr$^{-1}$ and 0.1 -- 100~Gyr, respectively (Fig.~1b). The XMP HI data (blue open circles, and red open square and circle; Fig.~1b) are roughly consistent with the position of many of the DGS sources (brown open triangles; Fig.~1a) and FIGGS galaxies (Roychowdhury et al. 2014) on the KS plot, and extend, towards higher $\Sigma _{\rm SFR}$ and $\Sigma _{\rm HI}$ values, the position of the LSB galaxies (light green open triangles; Fig.~1a). The XMP HI data are also roughly consistent with the range in HI depletion timescales observed for FIGGS galaxies at 400~pc resolution (Roychowdhury et al. 2015), with the average value for all HI-dominated regions in the HI Nearby Galaxy Survey (THINGS; Walter et al. 2008) 400~pc dataset (Roychowdhury et al. 2015), and with the range observed in dIrr galaxies (Elmegreen \& Hunter 2015).

Even considering the HI gas alone, XMPs are generally not consistent with the position of galaxies going through a merger process (purple solid line; Fig.~1b).

Although there is scatter due to uncertainties in the observations, a significant fraction of the XMPs already fall below the empirical KS star formation law (magenta solid line; Fig.~1b), implying that these XMPs are systems of low star-forming efficiency; they possess star formation surface densities too low for the values expected from the KS law for their gas masses. Nevertheless, these XMPs may still be consistent with the predictions of the Elmegreen (2015) (for log($\Sigma _{\rm gas}$/M$_{\odot}$ pc$^{-2}$) = 1, log($\Sigma _{\rm SFR}$/M$_{\odot}$ yr$^{-1}$ kpc$^{-2}$) = -2.8) and KMT+ (black solid line; Fig.~1b), models, in the HI-dominated regime, if the H$_2$ masses are not large (Sect.~4). 

In order to explore possible trends in the variation of the HI gas depletion timescales with the global properties, the XMPs have been plotted on the KS diagram (Fig.~2), where the sample was divided into sets of high (blue open circles) and low (red open circles) total HI mass, stellar mass, metallicity and optical source size. The limits between the high and low parameter range correspond, approximately, to the median values (except for the optical source size): log(M$_{\rm HI}$/M$_{\odot}$) = 8, log(M$_{\ast}$/M$_{\odot}$) = 7, 12+log(O/H) = 7.4, and $r_{\rm opt}$ = 3~kpc (which roughly corresponds to the 'dIrr'/'BCD' size limit; Sect.~2.1), respectively. No strong trend is observed for the metallicity (Fig.~2c) or size (Fig.~2d). In terms of the total HI mass (Fig.~2a) and stellar mass (Fig.~2b), it is observed that the more massive XMPs (blue open circles; Fig. 2a and 2b) tend to have the longest HI gas depletion timescales, and the least massive tend to have the shortest HI gas depletion timescales.



\setcounter{figure}{1}

\begin{figure*}	[h!]
\begin{center}

\includegraphics[width=7.5cm]{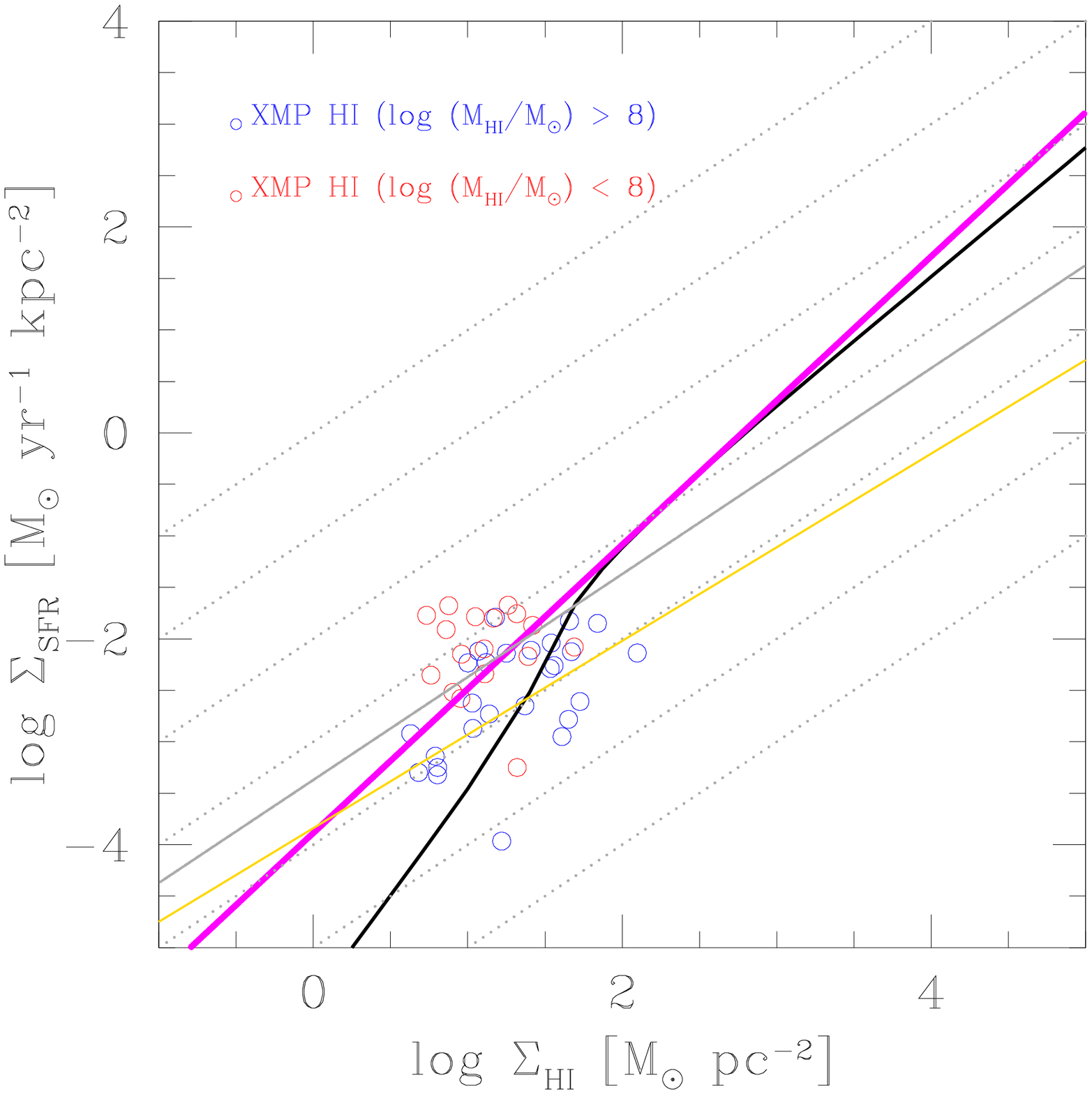}
(a)
\includegraphics[width=7.5cm]{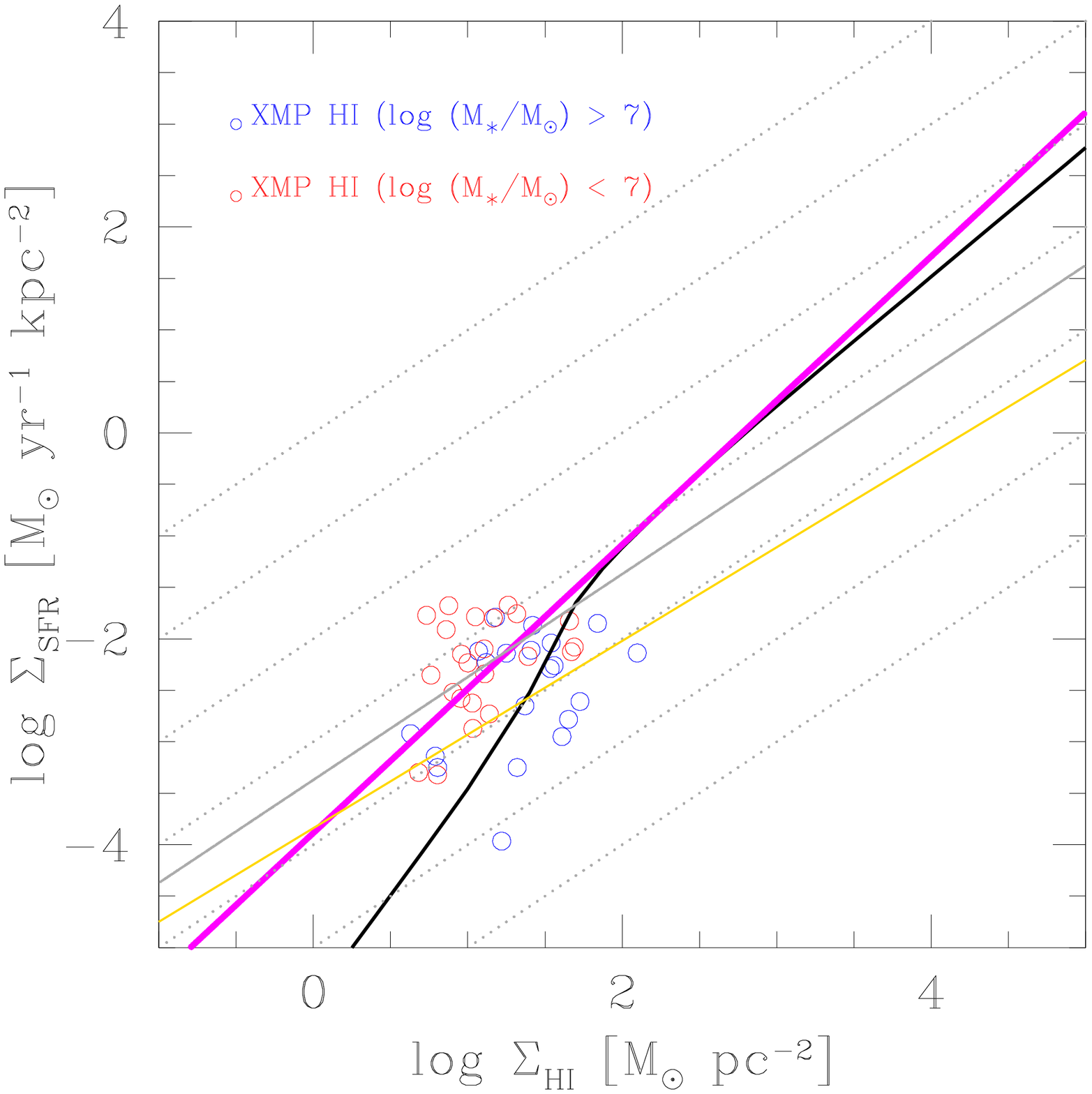}
(b)
\\
\includegraphics[width=7.5cm]{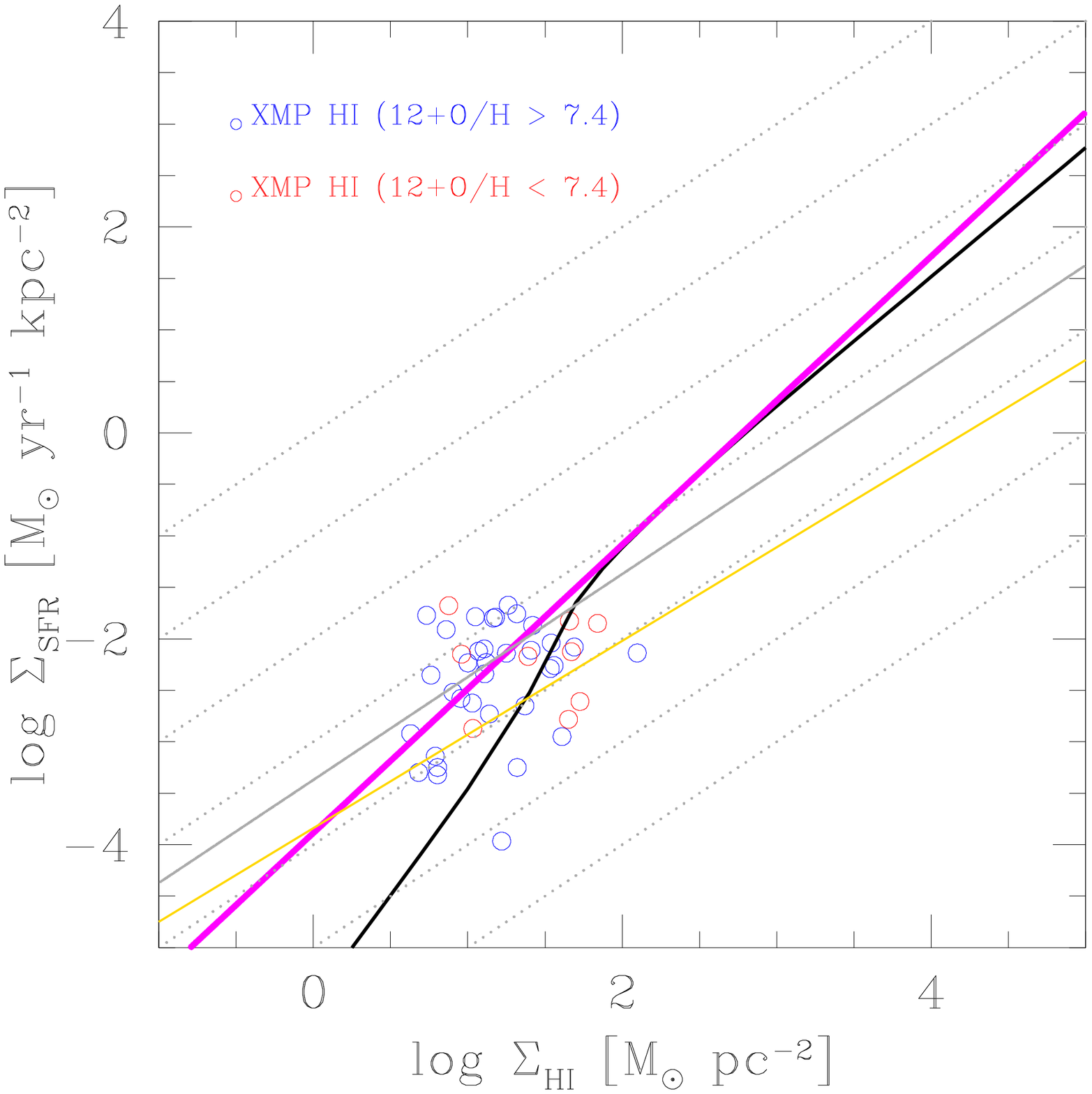}
(c)
\includegraphics[width=7.5cm]{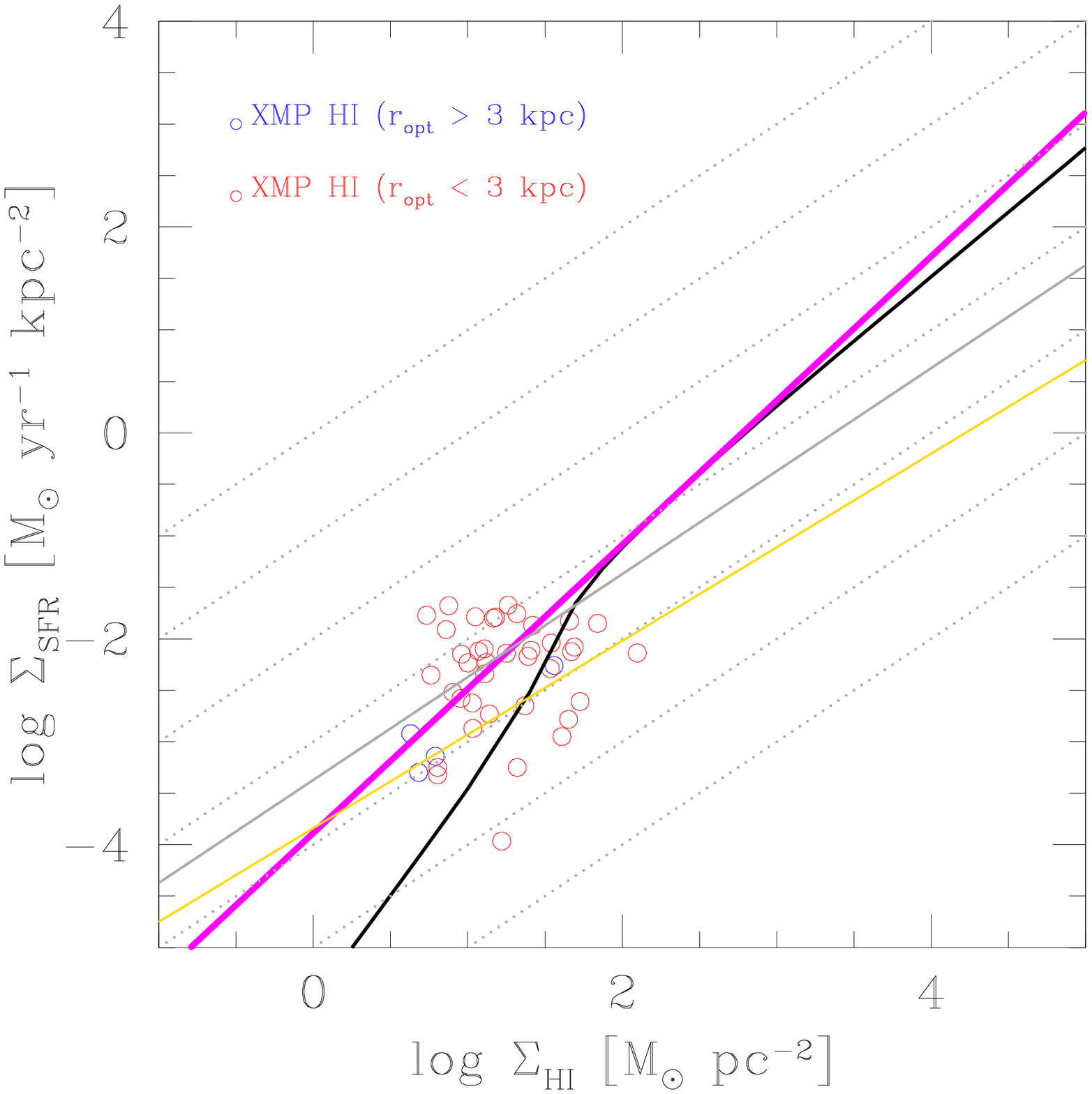}
(d)

\caption{The SFR surface density as a function of the HI gas surface density for the 43 sample XMPs, separated according to the (a) total HI mass (M$_{\rm HI}$/M$_{\odot}$, divided at 10$^8$), (b) stellar mass (M$_{\ast}$/M$_{\odot}$, divided at 10$^7$), (c) metallicity (12+O/H, divided at 7.4), and (d) optical source size ($r_{\rm opt}$, divided at 3~kpc). Red open circles correspond to the lower parameter range, and blue open circles to the higher parameter range. The lines are the same as in Fig.~1. See the caveat on low SFRs and low sSFRs in Section~2.}

\end{center}
\end{figure*}


Figure~3 contains scatter plots of the relations between the total HI mass (blue solid circles), the HI mass within the optical radius (red solid circles), the HI mass fraction (M$_{\rm HI}$/(M$_{\rm HI}$+M$_{\ast}$)), the HI star formation efficiency, the stellar mass, the SFR and the sSFR. All the plots show a large dispersion, although general trends may be pointed out. There is some tendency for the sSFR to be higher in the lower HI mass XMPs (Fig.~3a). The sSFR also shows a possible tendency to increase as the HI mass fraction of the HI mass within the optical radius (red solid circles) increases (Fig.~3b). However, the HI mass fraction of the total HI mass (blue solid circles) saturates to become of the order one, and independent of the sSFR, for sSFR $\gtrsim$ 10$^{-9}$ yr$^{-1}$ (Fig.~3b). It is further observed that the smaller the HI mass, the faster the XMP consumes its HI gas (Fig.~3c). The SFR also shows a possible tendency to increase with HI mass (Fig.~3d). However, disregarding the lower SFR points (see caveat on low SFRs in Section~2), for a fixed SFR, the mass of the HI reservoir can vary by approximately three orders of magnitude (Fig.~3d).




\setcounter{figure}{2}
\begin{figure*}	
\begin{center}

\includegraphics[width=7.5cm]{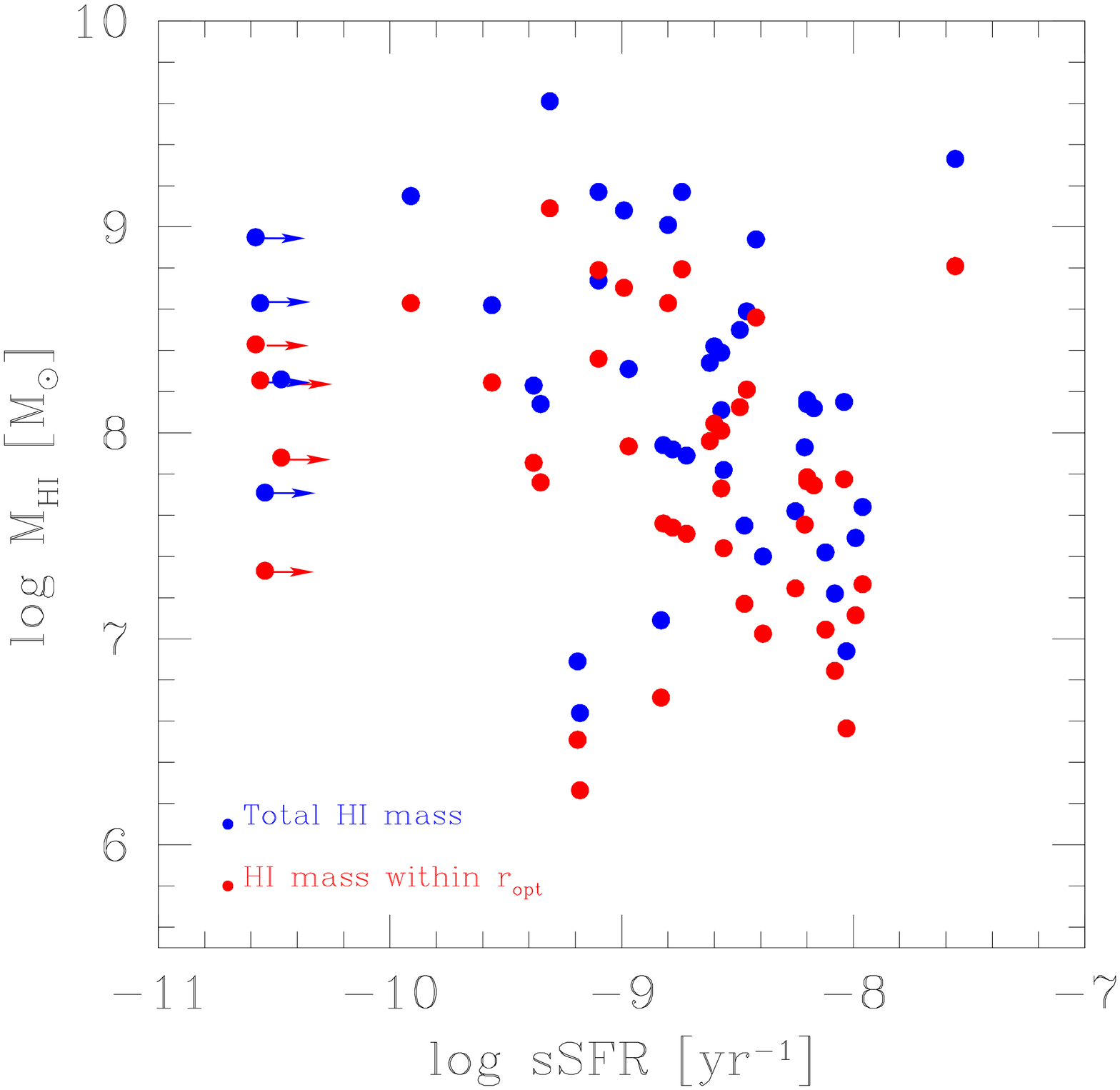}
(a)
\includegraphics[width=7.5cm]{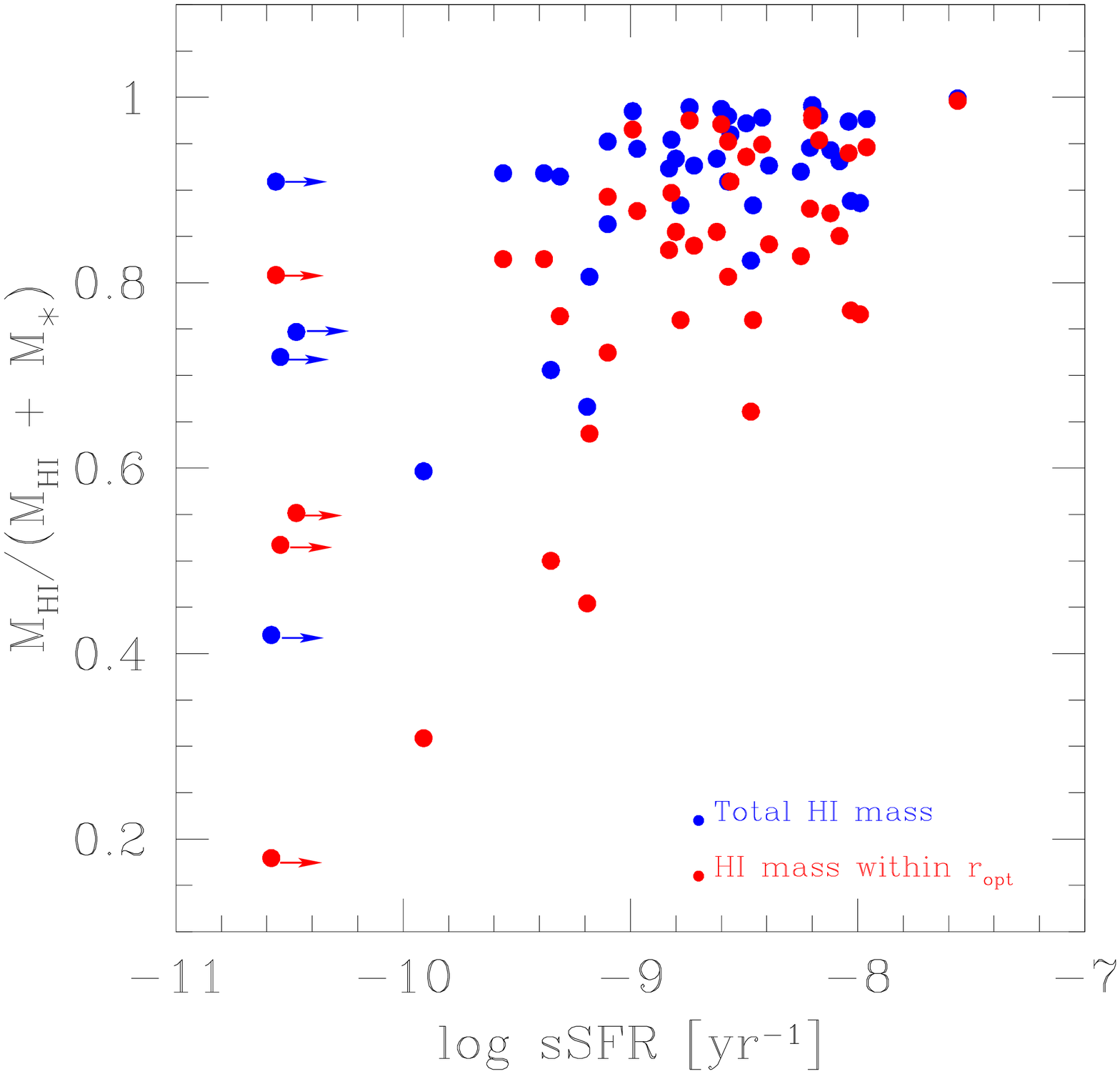}
(b)
\includegraphics[width=7.5cm]{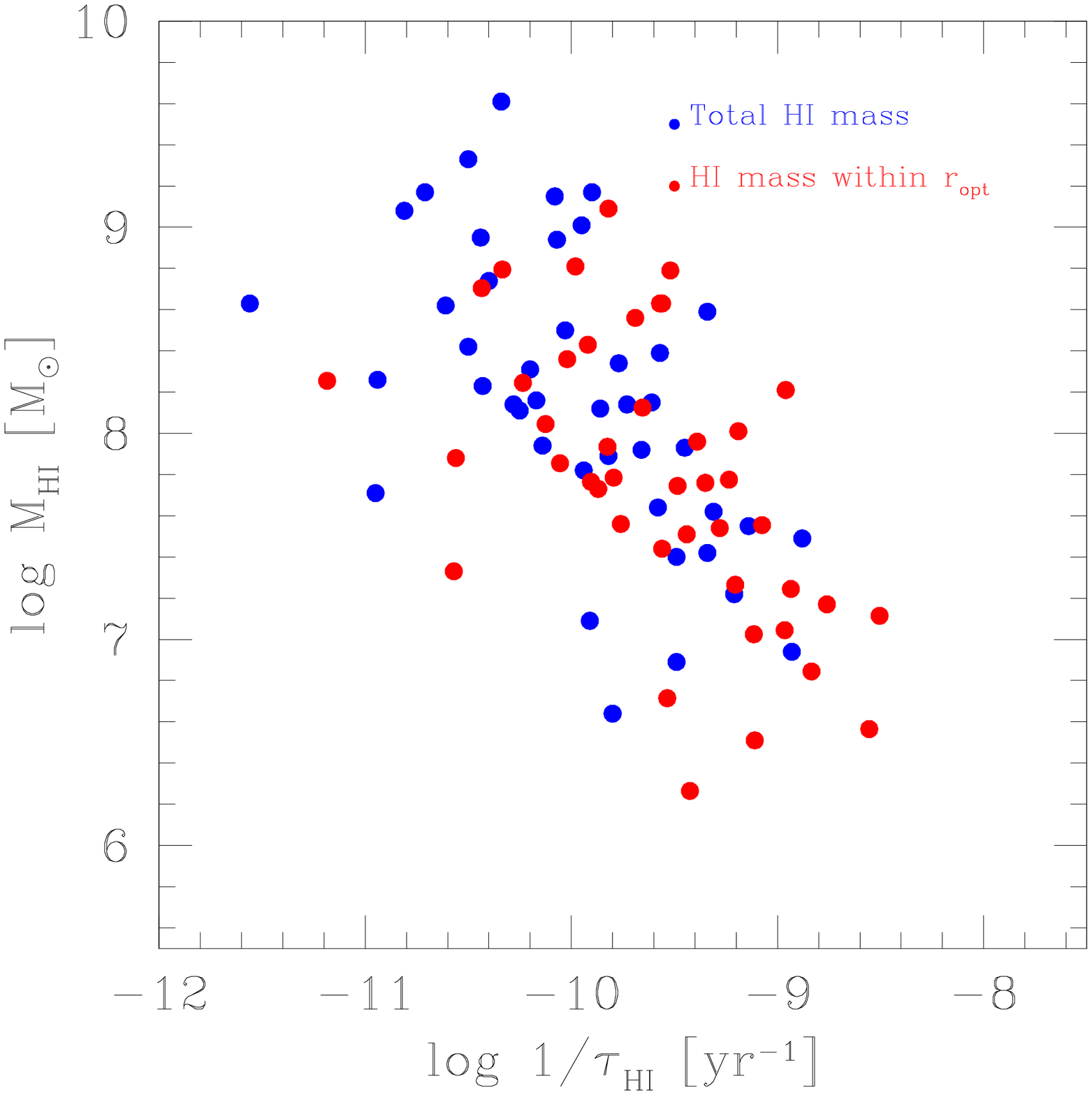}
(c)
\includegraphics[width=7.5cm]{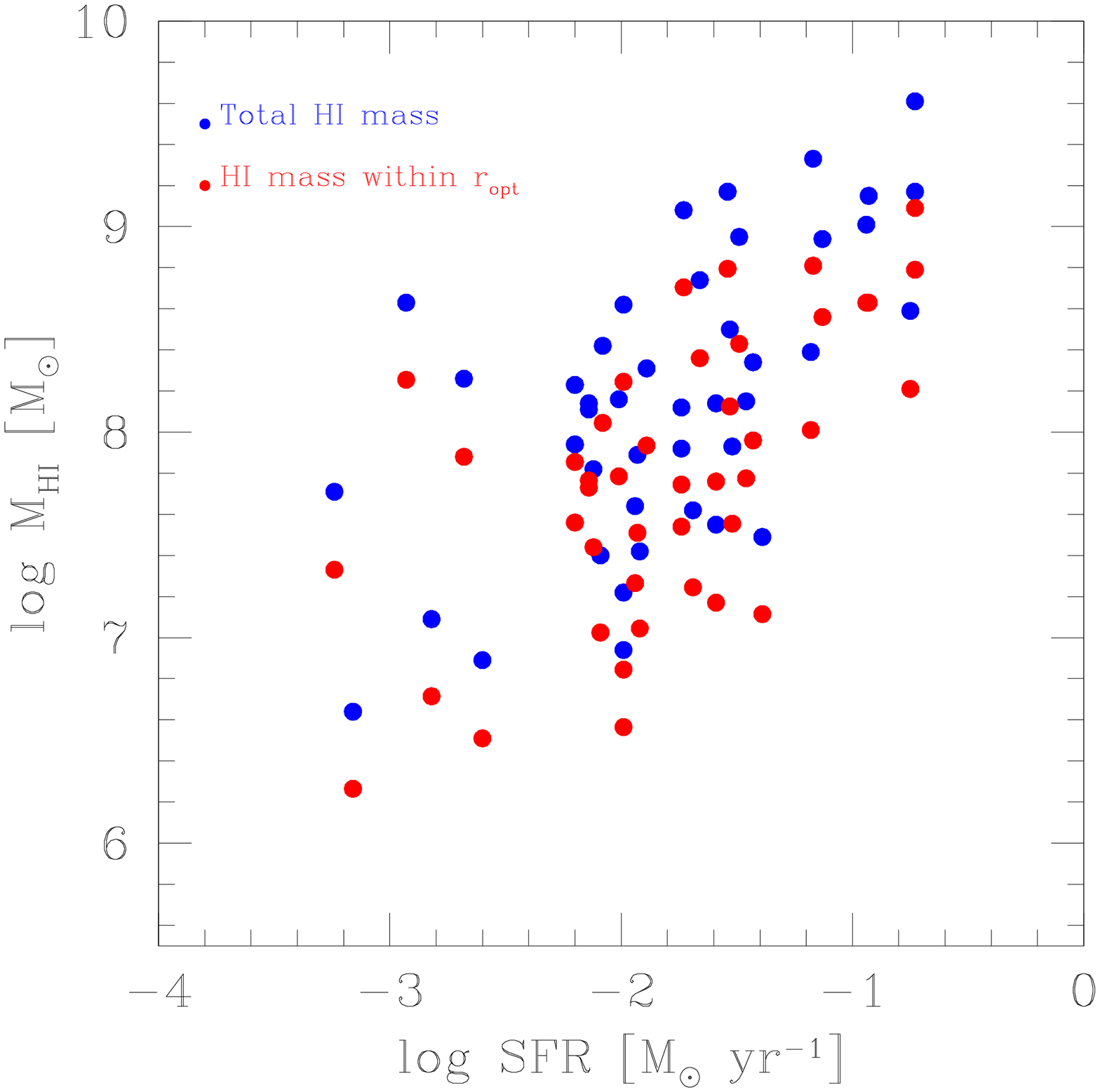}
(d)

\caption{Relations between integrated physical parameters for the 43 sample XMPs. Blue and red solid circles correspond to using, for the estimates, the total HI mass and the HI mass within the optical radius, respectively. (a) The HI mass as a function of the specific SFR (sSFR $\equiv $ SFR/M$_{\ast}$). (b) The HI mass fraction (M$_{\rm HI}$/(M$_{\rm HI}$ + M$_{\ast}$)), as a function of the sSFR. (c) The HI mass as a function of the HI star formation efficiency (SFE$_{\rm HI} \equiv$ 1/$\tau_{\rm HI}$). (d) The HI mass as a function of the SFR. See the caveat on low SFRs and low sSFRs in Section~2.}

\end{center}
\end{figure*}




\setcounter{figure}{3}
\begin{figure*}	
\begin{center}

\includegraphics[width=7.5cm]{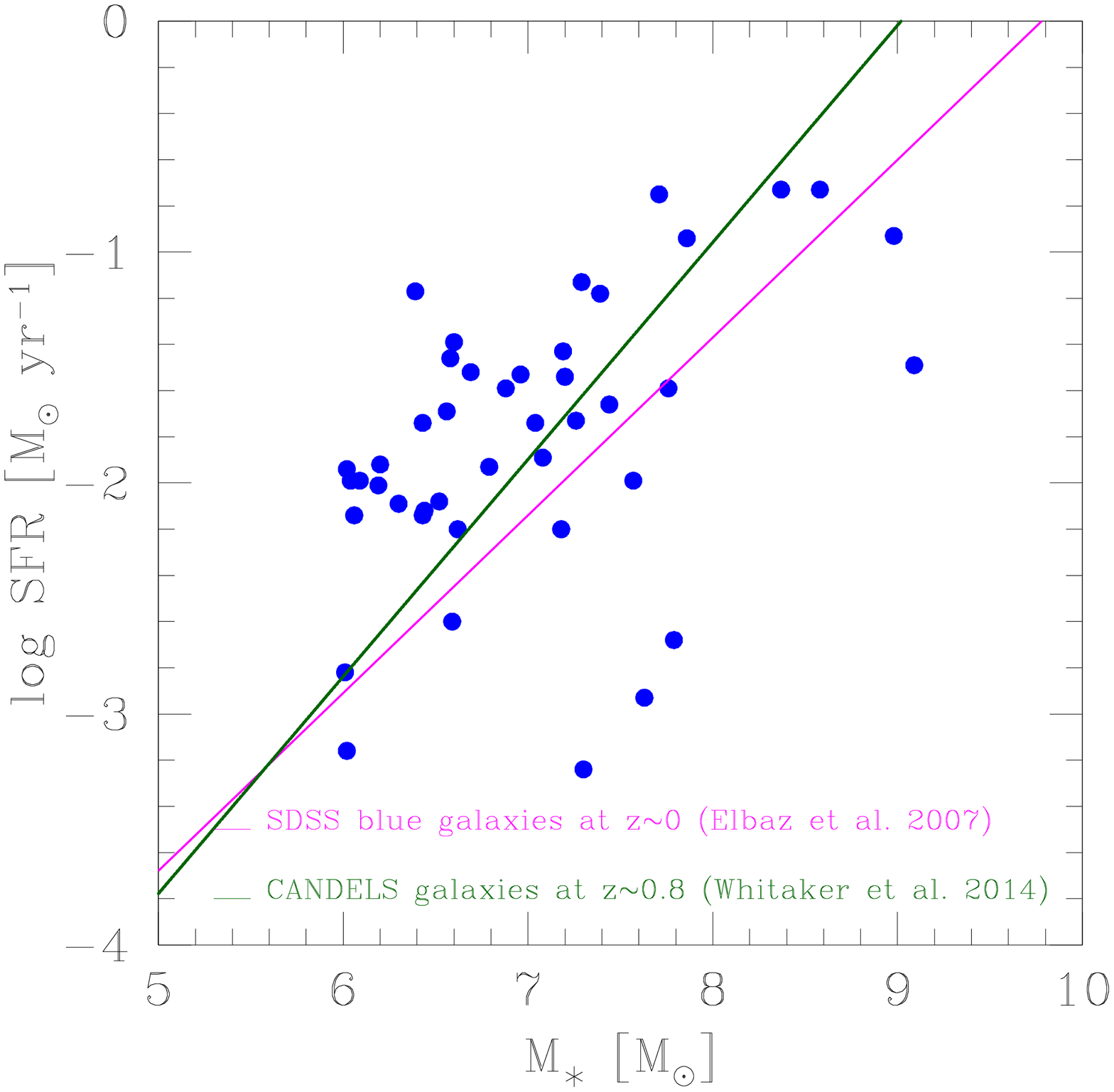}
(a)
\includegraphics[width=7.5cm]{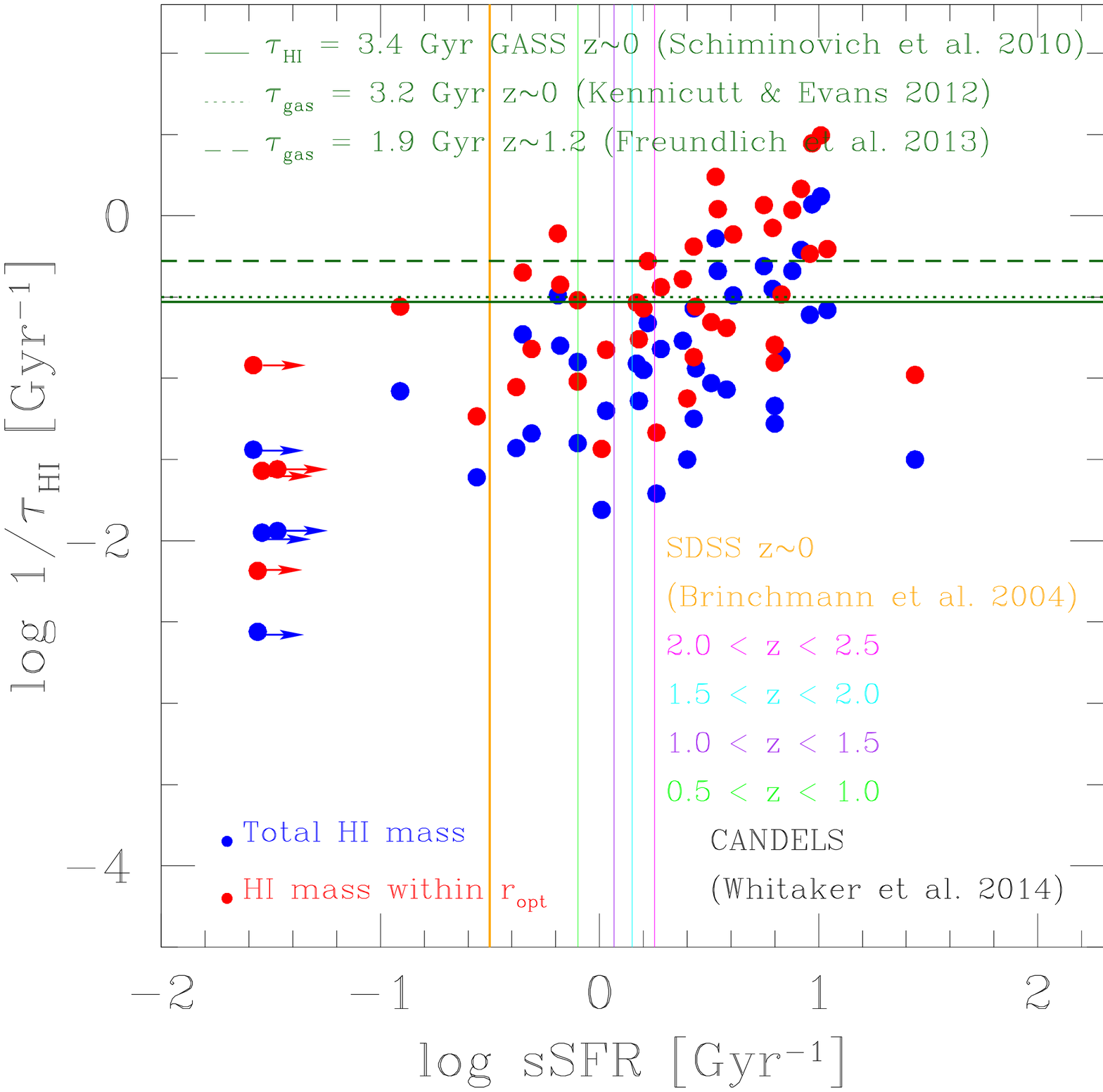}
(b)
\caption{Relations between integrated physical parameters for the 43 sample XMPs. (a) The SFR as a function of the stellar mass. The plot contains extrapolated (to low stellar mass) observed relations for SDSS blue galaxies at $z \simeq $ 0 (magenta solid line; Elbaz et al. 2007) and CANDELS galaxies at $z \simeq $ 0.8 (dark green solid line; Whitaker et al. 2014). (b) The HI star formation efficiency (SFE$_{\rm HI} \equiv$ 1/$\tau_{\rm HI}$) as a function of the sSFR (sSFR $\equiv $ SFR/M$_{\ast}$). The plot includes an average HI gas depletion timescale observed for GASS galaxies at $z \simeq $ 0 ($\tau _{\rm HI}$ = 3.4~Gyr; dark green solid line; Shiminovich et al. 2010), the total gas depletion timescale corresponding to the KS prediction at $z \simeq $ 0 ($\tau _{\rm gas}$ = 3.2~Gyr; dark green dotted line; Kennicutt \& Evans 2012), and the total gas depletion timescale corresponding to the observed KS law at $z \simeq $ 1.2 ($\tau _{\rm gas}$ = 1.9~Gyr; dark green dashed line; Freundlich et al. 2013). In addition, the figure also includes, as a reference, vertical lines with a linear extrapolation (to low stellar mass) of the observed sSFR values for SDSS galaxies at $z \simeq $ 0 (orange solid line; Brinchmann et al. 2004) and CANDELS galaxies at 0.5 $ < z < $ 2.5 (magenta, cyan, purple and light green solid line; Whitaker et al. 2014). See the caveat on low SFRs and low sSFRs in Section~2.}

\end{center}
\end{figure*}


Figure~4a shows the SFR as a function of the stellar mass, and contains two observed relations extrapolated down to low stellar masses: the so-called main sequence of star formation for the Cosmic Assembly Near-Infrared Deep Extragalactic Legacy Survey (CANDELS) at $z \simeq $ 0.8 (broken power law, fit to lower stellar masses; dark green solid line; Whitaker et al. 2014) and for the SDSS blue galaxies at $z \simeq $ 0 (magenta solid line; Elbaz et al. 2007). When compared to the extrapolated predictions, a significant fraction of the XMPs show elevated SFRs relative to their stellar masses at $z \simeq $ 0 (magenta solid line; Fig.~4a); they appear to possess sSFRs similar to galaxies at $z \simeq $ 1 (dark green solid line; Fig.~4a). The fact that XMPs possess sSFRs similar to higher redshift galaxies is further evidenced in Figure~4b, where the HI star formation efficiency is plotted against the sSFR. The figure includes the sSFRs expected from the linear extrapolation to low stellar masses of CANDELS galaxies at 0.5 $ < z < $ 2.5 (magenta, cyan, purple and light green solid lines; Whitaker et al. 2014) and SDSS galaxies at $z \simeq $ 0 (orange solid line; Brinchmann et al. 2004). For reference, Figure~4b also includes an average HI gas depletion timescale for GALEX-Arecibo-SDSS Survey (GASS) galaxies at $z \simeq $ 0 ($\tau _{\rm HI}$ = 3.4~Gyr; dark green solid line; Schiminovich et al. 2010), the total gas depletion timescale corresponding to the KS prediction at $z \simeq $ 0 ($\tau _{\rm gas}$ = 3.2~Gyr; dark green dotted line; Kennicutt \& Evans 2012), and the total gas depletion timescale corresponding to the observed KS law at $z \simeq $ 1.2 ($\tau _{\rm gas}$ = 1.9~Gyr; dark green dashed line; Freundlich et al. 2013). It is to be noted that XMPs tend to have sSFRs in excess of the typical values observed in the local Universe (orange solid line; Fig.~4b). It is also observed that, generally, the higher the sSFR (and the higher the SFR surface densities), the faster the XMP depletes its HI gas (Fig.~4b). 



Figure~4b further demonstrates that the HI gas depletion timescales, i.e., $\tau _{\rm HI}$, can be almost an order of magnitude longer than the timescale to form the observed stellar mass at the current SFR, i.e., 1/sSFR.  However, the HI gas depletion timescales inferred from the HI mass are only conservative upper limits if galaxy winds are important. A significant part of the (HI) gas may be expelled into the circum-galactic medium, so that the 'effective' HI gas depletion timescales ($\tau _{\rm HI}^{\rm eff}$) may be much shorter:


\begin{equation}
\tau _{\rm HI}^{\rm eff} = \tau _{\rm HI}/(1+W-R),
\end {equation}

\noindent where $R$ is the fraction of the HI gas returned to the interstellar medium by stellar winds and supernova explosions, and $W$ is the mass loading factor (e.g., S\'anchez Almeida et al. 2014). $W$ is the ratio between the SFR and the mass loss rate through winds, and can be very large in dwarf galaxies; for example, Dayal, Ferrara \& Dunlop (2013) use $W$ in excess of 10 to reproduce the mass-metallicity-SFR relation. For typical values of $R$ = 0.2 and $W$ = 5, the 'effective' HI gas depletion timescales are shortened by a factor of approximately six, increasing the 'effective' HI star formation efficiencies to become of the same order as the specific star formation rates, i.e., SFE$_{\rm HI}^{\rm eff} \equiv$ 1/$\tau _{\rm HI}^{\rm eff} \simeq$ sSFR. The one-to-one relation between the SFE$_{\rm HI}^{\rm eff}$ and sSFR is shown in Figure~5 (grey solid line), which is similar to Figure~4b, but where the specific star formation rate has been compared with the 'effective' HI star formation efficiency (red solid circles).



\setcounter{figure}{4}
\begin{figure}	
\begin{center}

\includegraphics[width=7.5cm]{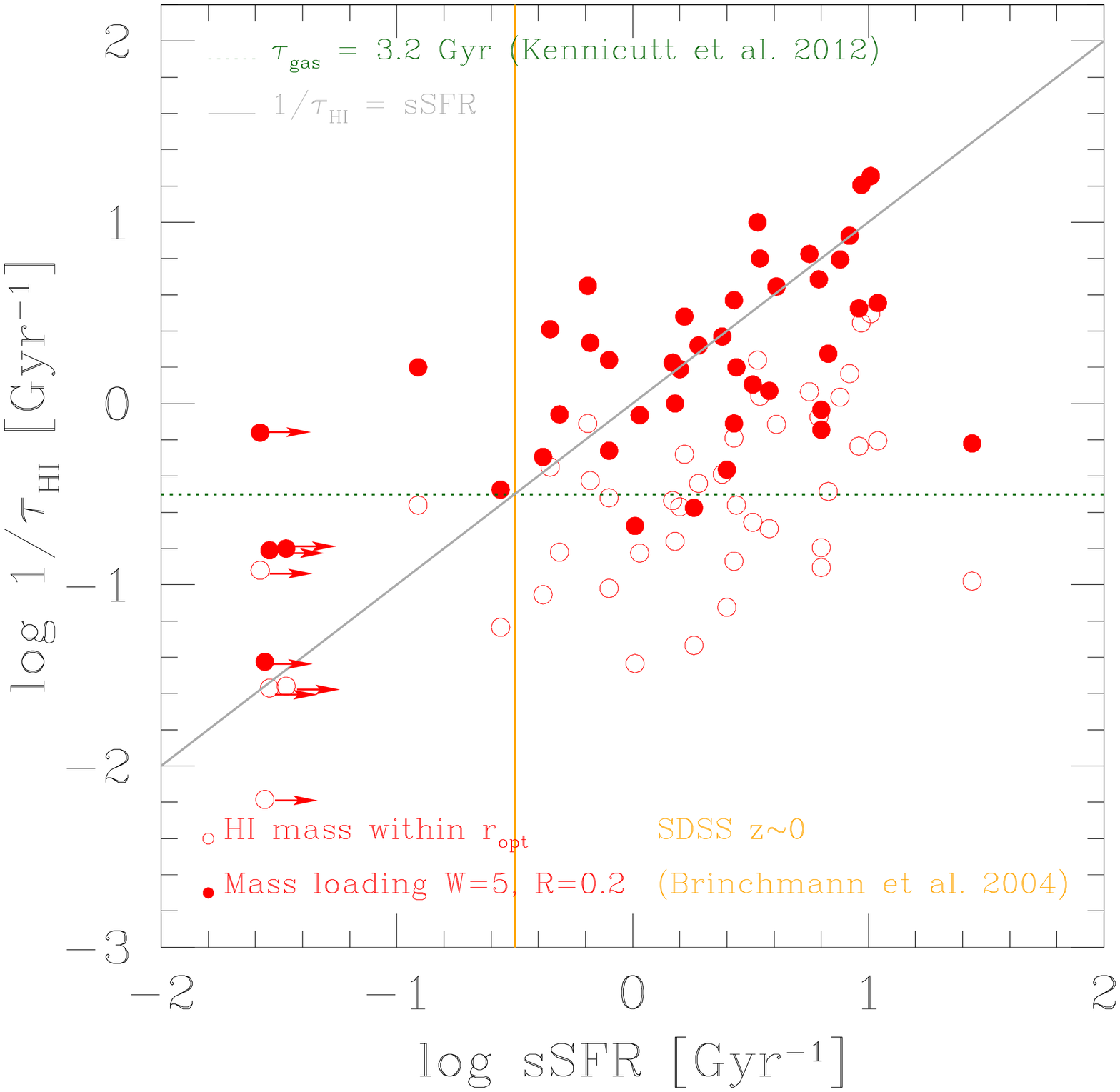}

\caption{The HI star formation efficiency (SFE$_{\rm HI} \equiv$ 1/$\tau_{\rm HI}$) as a function of the sSFR (sSFR $\equiv $ SFR/M$_{\ast}$) for the HI mass within the optical galaxy (red open circles), and when galaxy winds, with mass loading factors of $W$ = 5, are used to compute the 'effective' HI gas depletion timescale (red solid circles). When winds are included, the HI star formation efficicency and sSFR become of the same order (grey solid line). The plot also includes the total gas depletion timescale corresponding to the KS prediction at $z \simeq $ 0 ($\tau _{\rm gas}$ = 3.2~Gyr; dark green dotted line; Kennicutt \& Evans 2012), and a linear extrapolation (to low stellar mass) of the observed sSFR values for SDSS galaxies at $z \simeq $ 0 (orange solid line; Brinchmann et al. 2004). See the caveat on low SFRs and low sSFRs in Section~2.}
\end{center}
\end{figure}


The above results highlight the fact that (extreme) low metallicity alone can not account for the plethora of XMP dust and gas properties, and, therefore, star-forming efficiency (e.g., Royhowdhury et al. 2015); star formation triggering, timescales, gas accretion, mass ratios, high gas mass and density, and mass loading are likely playing significant and interdependent roles (e.g., Hunt et al. 2014; Rubio et al. 2015).


\section{Significance of the Molecular Mass}

Because there are no available H$_2$ measurements for the XMPs, with the exception of upper limits to three sources (Cormier et al. 2014), molecular gas masses for the 43 sample XMP galaxies have been derived (Table~2) using published empirical scaling relations. For the scaling between metallicity, M$_{\rm HI}$, SFR and M$_{\rm H_2}$, various recipes, existing in literature, have been considered:

\begin{enumerate}

\item{M$_{\rm H_2}$/M$_{\rm HI}$ = 6 (Shi et al. 2014);}

\item{log(M$_{\rm H_2}$) = 1.2 log(M$_{\rm HI}$) - 1.5 (12+log(O/H) - 8.7) - 2.2 (Amor\'\i n et al. 2015);}

\item{M$_{\rm H_2}$ = SFR $\times$ $\tau _{\rm H_2}$, where $\tau _{\rm H_2}$ = 2.35~Gyr (the 'universal' H$_2$ depletion timescale; Bigiel et al. 2008, 2010);}

\item{M$_{\rm H_2}$ = SFR $\times$ $\tau _{\rm H_2}$, where $\tau _{\rm H_2}$ = 1~Gyr (the 'short' H$_2$ depletion timescale).} 

\end{enumerate}

The different derived molecular gas masses (Table~2), as a function of the measured HI masses, are included in Figure~6a. In Figure~6b and 6c, the KS plot of Figure~1b is reproduced with the 43 sample XMP HI data points (blue open circles), and the derived total gas surface densities contained in Table~2 (crosses). Using the SFR and HI data to predict the H$_2$ content recovers a broad range (2 -- 3 orders of magnitude) in derived molecular gas masses (Fig.~6).




\setcounter{figure}{5}

\begin{figure*}	
\begin{center}

\includegraphics[width=7.5cm]{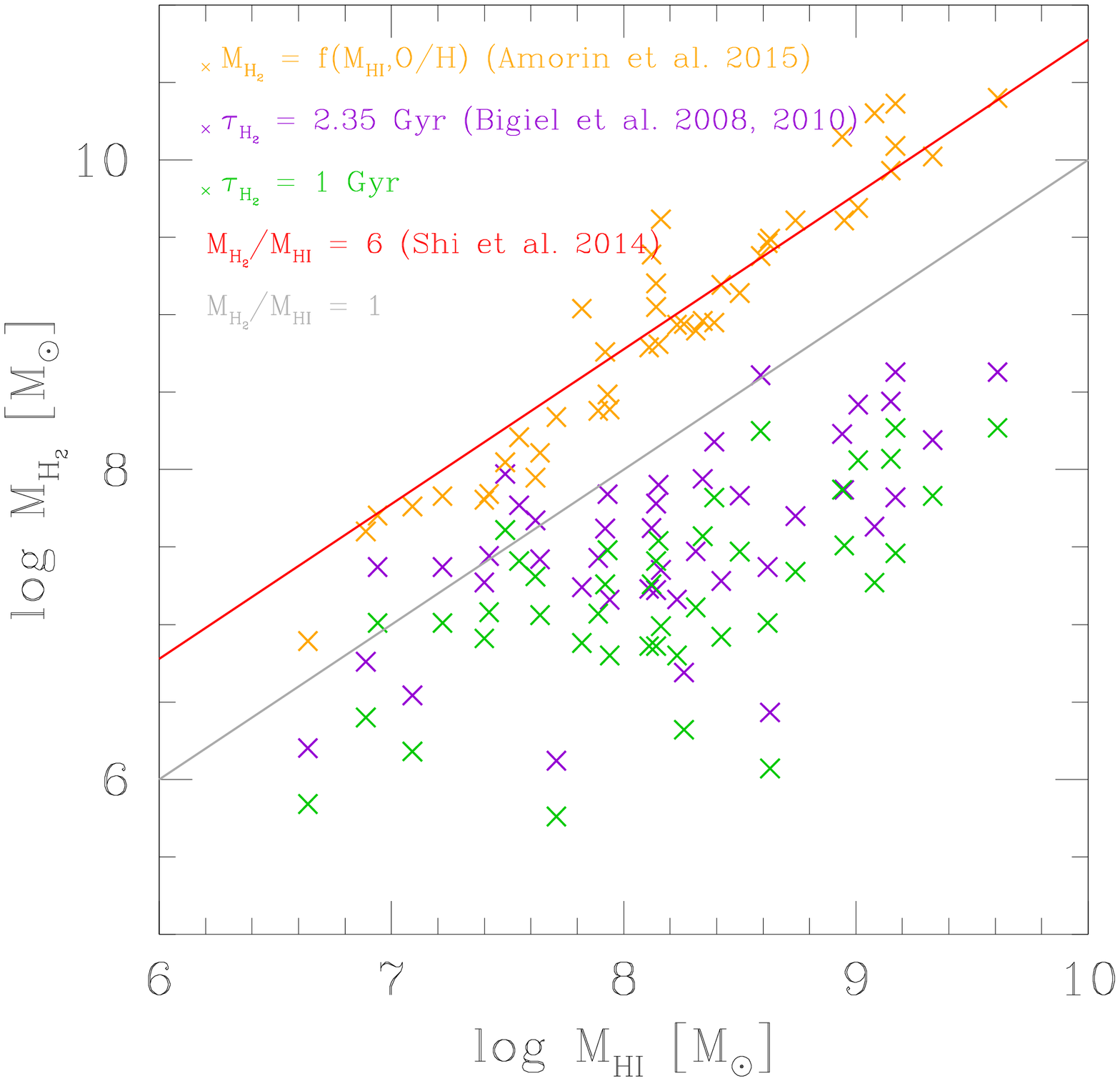}
(a)
\\
\includegraphics[width=7.5cm]{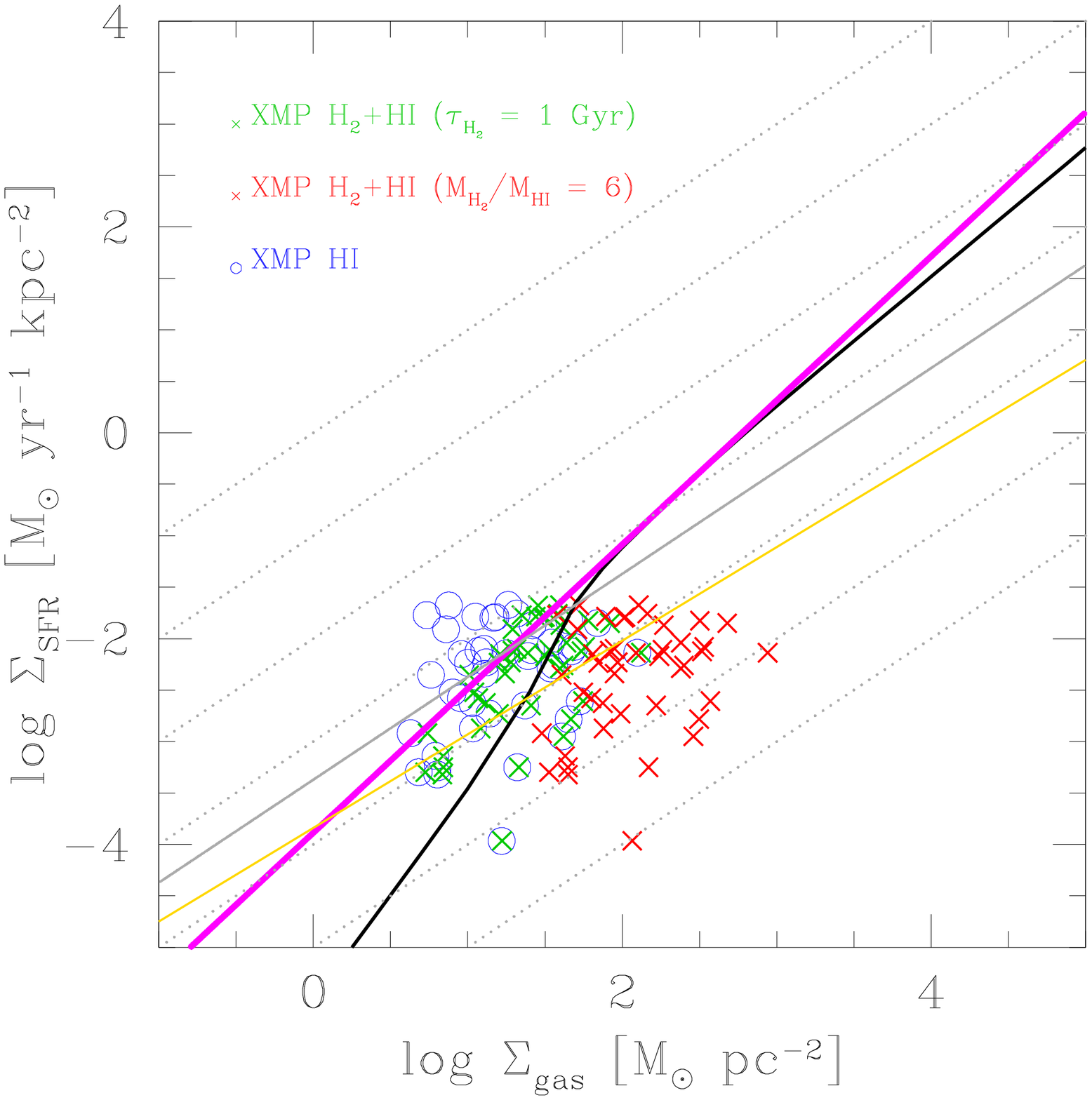}
(b)
\includegraphics[width=7.5cm]{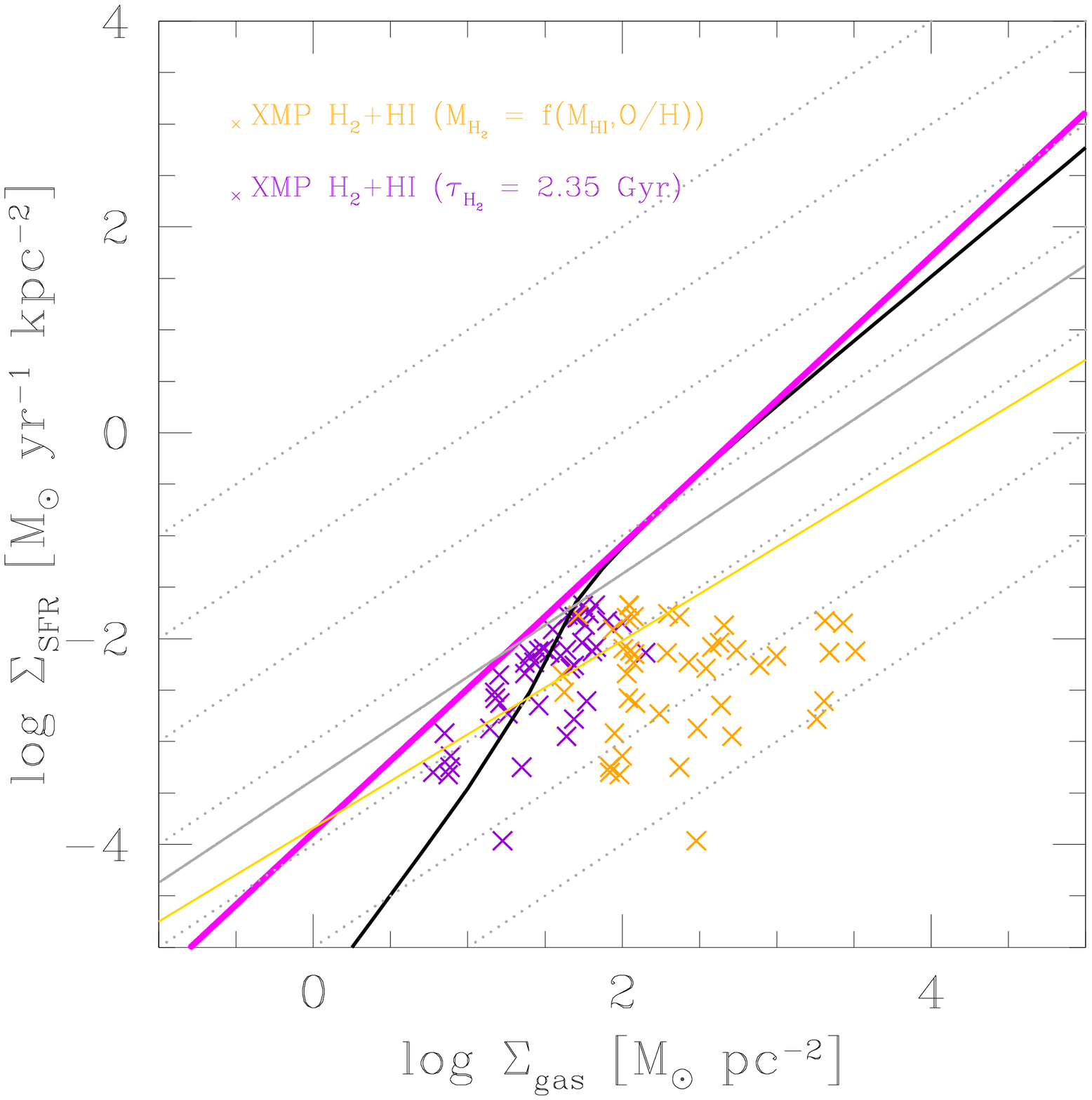}
(c)

\caption{(a) The derived H$_2$ mass as a function of the observed HI mass (within the optical radius) or SFR, for several empirical scaling relations. The grey solid line corresponds to the line of equal H$_2$ and HI mass. The red solid line corresponds to the derived H$_2$ masses assuming the Shi et al. (2014) relation. The purple, green and yellow crosses correspond to H$_2$ masses derived assuming the 'universal' ($\tau _{\rm H_2} = $ 2.35~Gyr; Bigiel et al. 2008, 2010) and 'short' ($\tau _{\rm H_2} = $ 1~Gyr) H$_2$ gas depletion timescale, and the Amor\'\i n et al. (2015) relation, respectively. (b) and (c) The SFR surface density as a function of the total (crosses) or HI (open circles) gas surface density for the 43 sample XMPs. Purple, green, red and yellow crosses correspond to total gas surface densities estimated as assigned above. The lines are the same as in Fig.~1. See the caveat on low SFRs and low sSFRs in Section~2.}

\end{center}
\end{figure*}


It has been shown that gas depletion timescales shorten with mass (Fig.~3c; Leroy et al. 2005; Hunt et al. 2015); smaller, less massive dwarf galaxies tend to deplete their gas faster. Because the sample sources are generally low mass, then $\tau _{\rm H_2} $ = 1~Gyr may be an appropriate value for the H$_2$ gas depletion timescale in XMPs (green crosses; Fig.~6a and 6b). In this case, the derived H$_2$ masses are generally low, with M$_{\rm H_2} \lesssim $ M$_{\rm HI}$ (green crosses; Fig.~6a). The resulting $\Sigma _{\rm gas}$ values (green crosses; Fig.~6b) are roughly consistent with the predictions of the KMT+ model (black solid line; Fig.~6b).

However, the molecular gas may be far more significant, as implied by Sextans A and ESO 146-G14 (red solid squares and circles, respectively; Fig.~1a), some DGS sources (brown solid triangles; Fig.~1a), and dwarf galaxies (Amor\'\i n et al. 2015), and as suggested by recent [CII] observations of metal-poor environments (e.g., Madden 2000; Cormier et al. 2014; Cormier et al. 2015). If the molecular gas is over five times more abundant than the HI gas (Fig.~6a), then a significant fraction of the XMPs may possess large total gas surface densities (red and orange crosses; Fig.~6b and 6c), accompanied by extremely low star formation efficiencies ($\tau _{\rm gas} \gtrsim $ 10~Gyr). These (total and H$_2$) gas depletion timescales can be compared to the average values provided for the resolved observations of Sextans A and ESO 146-G14 ($\tau _{\rm H_2} \simeq $ 60~Gyr; Shi et al. 2014), subparsec-scale values in local spirals ($\tau _{\rm H_2} \simeq $ 2~Gyr; Bigiel et al. 2008; Leroy et al. 2008), and integrated values for dwarf galaxies ($\tau _{\rm H_2} \simeq $ 0.6~Gyr and $\tau _{\rm gas} \simeq $ 5~Gyr; Cormier et al. 2014; see also Amor\'\i n et al. 2015). If, indeed, such large amounts of (as yet undetected) H$_2$ gas exist in XMPs, associated with large reservoirs of HI gas and an extremely low star-formation efficiency, then that would necessarily imply a mechanism that efficiently converts HI gas into H$_2$, while greatly suppressing the star formation in these galaxies. 

\section{Discussion}

In light of the available auxiliary and XMP data, the location of a source on the KS plot may be dependent on the role of the HI gas in the star formation process, and, in some cases, the position on the KS diagram may be transient.

It is known that merging/starburst galaxies do not follow the 'classical' KS law (Kennicutt \& Evans 2012), but a displaced star formation law; mergers/starbursts exhibit an enhancement of the SFR for their total gas (purple solid line; Fig.~1a), which is rapidly depleted by cooling and compression to form stars (e.g., Daddi et al. 2010). Several dwarf galaxies (Cormier et al. 2014) occupy this region of the KS plot, and may represent merging systems in an advanced stage of interaction (Fig.~1a). 

Galaxies and galaxy regions (e.g., centers of galaxies) that follow the KS law are generally sources in which the star formation mode is secular or long-lasting, i.e., the observed HI reservoir is involved in the star formation process, and is being appropriately depleted (e.g., Kennicutt 1998; Daddi et al. 2010; Kennicutt \& Evans 2012). 

At the other extreme, there are sources with longer total gas depletion timescales than predicted by the KS law (Fig.~1a and 1b), which are suggested to be either interacting sources/sources with companions, or sources whose global HI reservoir is involved, in a different manner, in the star formation process (e.g., Cormier et al. 2014). In the former case, the proximity with a close companion may cause an overestimation of the HI mass (within the large beam) of the main source. Alternatively, the interaction/presence of a companion may provide the main source with fresh HI gas, but this gas has not yet begun forming stars at a high rate. In a more advanced stage of the interaction, the source may move to a region of the KS plot corresponding to larger SFR surface densities. 

Although some notable XMPs possess companions (e.g., IZw 18 and SBS 0335-052; see Filho et al. 2015 and references therein for details; see also the Appendix), the majority appear to be relatively isolated (Filho et al. 2015). Therefore, the latter scenario, that of a different participation of the HI reservoir in the star formation process, relative to main spiral galaxy disks, is the scenario favoured for the XMPs (and perhaps in some other 'isolated' metal-poor dwarf galaxies). In the particular case of low metallicity conditions, efficient gas cooling and H$_2$ formation, due to the lack of metals and dust, are strongly constrained. It has also been hypothesized that, in low metallicity environments, star formation may proceed even before most of the atomic gas is converted into molecular form (e.g., Glover \& Clark 2012a; Krumholz 2012). However, dust grains may be present in some XMP star-forming regions, from previous star formation episodes or from contamination of neighbouring regions, partly catalyzing H$_2$ formation and shielding the molecular gas from the ionizing radiation. This appears to be the case for Sextans A and ESO 146-G14 (Shi et al. 2014), which show, on star-forming region scales, significant amounts of dust, and, therefore, dust-determined H$_2$ masses, associated with a low star formation efficiency (Fig.~1a). Large quantities of 'dark' H$_2$ mass are also inferred from the large [CII]-to-CO line fluxes observed, on kiloparcsec scales, in several low metallicity dwarf galaxies (Cormier et al. 2014, 2015), where the [CII] acts as a gas coolant (e.g., Wolfire, Hollenbach \& McKee 2010; Cormier et al. 2015). In the case of the three XMPs which possess upper limit CO determinations, these provide [CII]/CO ratios that can be several hundred thousand (IZw 18) to less than several thousand (SBS 0335-052; Cormier et al. 2014, 2015). It is to be noted, however, that the [CII] may also be tracing the HI gas, which is known to be significant in XMPs (Filho et al. 2013).

The existence of such large amounts of gas in XMPs would imply the suppression of star formation. Four, potentially interdependent, factors are proposed to have relevant impact on the observed low star formation efficiency in XMPs, compared to the main disks of spiral galaxies:

\begin{enumerate}

\item{The entire HI reservoir may participate in the star formation process (Fig.~3c), but its metal-poor nature (Filho et al. 2013) may inevitably lead to a quashing of the global star formation rate with respect to the KS law, such as the type of effect modelled by Krumholz (2013).}

\item{Most of the HI reservoir is inert, and only a small fraction participates in the star formation process when forced by an external triggering event. This is consistent with the large range in HI mass observed for a fixed SFR (Fig.~3d), and with the detection of metallicity drops associated with starbursts in XMPs (S\'anchez Almeida et al. 2013, 2015). As is observed in many BCDs (e.g., Adamo, \"Ostlin \& Zackrisson et al. 2011), most of the star formation activity occurs in massive clumps. As such, it has been suggested that the star formation in these regions is fed by the accretion of metal-poor HI gas via cosmological cold gas accretion (e.g., \"Ostlin et al. 2001; Ekta \& Chengalur 2010; S\'anchez Almeida et al. 2013, 2014, 2015; Filho et al. 2013, 2015). Cold gas accretion is predicted to occur by numerical simulations, and is considered a relevant driver of star formation, even at low redshift (e.g., Birnboim \& Dekel 2003; Kere\v {s} et al. 2005; Dekel \& Birnboim 2006; Brooks et al. 2009). Star formation often occurs before the infalling gas is mixed with the pre-existing gas, and starbursts ensue in metal-poor environments (Ceverino et al. 2015).}

\item{The time lag to trigger star formation upon the arrival of fresh HI gas may also play a significant role; XMPs with extremely low star formation efficiencies (Fig.~1b) may be sources that have just received a fresh supply of HI gas, but have not had time to appropriately adjust their SFR (e.g., Cormier et al. 2014).}

\item{Galaxy winds are expected to be important in dwarf galaxies (e.g., De Young \& Gallagher 1990; Silich \& Tenorio-Tagle 2001; S\'anchez Almeida et al. 2014). For typical mass loading factors, more than 80\% of the gas may be returned back to the circum-galactic medium (Fig.~5), contributing to the lowering of the star formation efficiency.}

\end{enumerate}

\noindent Therefore, in XMPs, the global HI mass is likely not a good tracer of the H$_2$ mass nor of the SFR (as given by the KS law).

It is becoming increasingly clear that star formation triggering, timescales, gas accretion, mass ratios, high gas mass and density, and mass loading may have a significant impact on the (local and global) efficiency of the star formation process in low metallicity environments, perhaps more so than the metallicity itself (e.g., Hunt et al. 2014; Roychowdhury et al. 2015; Rubio et al. 2015).

Whether, in XMPs, star formation is locally efficient (e.g., S\'anchez Almeida et al. 2015) or not (e.g., Shi et al. 2014; Rubio et al. 2015), the net global effect is that XMPs are HI-dominated, high specific star-forming (sSFR $\gtrsim $ 10$^{-10}$~yr$^{-1}$), low star formation efficiency (SFE$_{\rm gas} \lesssim$ 10$^{-9}$~yr$^{-1}$) systems.  
 
\section{Conclusions}



Traditional H$_2$ mass tracers, such as CO and dust, and, more recently, the use of the [CII] line, appear to provide, in metal-poor environments, discrepant estimations of the H$_2$ mass (Sect.~1). This stems from the unique conditions for star formation that are available in such environments. In order to probe these unique environments, we have explored the low surface density and SFR regime of the KS diagram, comparing with empirical and theoretical star formation laws, as well as with observational data.

The observational data employed in the analysis consist of auxiliary data from literature for dwarf and LSB galaxies (Sect.~2.2), plus data for a 43 source sample of XMPs (Sect.~2.1). For the latter dataset, the observational parameters have been compiled from published HI data, from SDSS or DSS-II images, and SFR data from the MPA-JHU. The relevant quantities have been derived from these parameters without performing inclination corrections (due to the irregular optical morphology of the XMPs), and assuming a very simplified model, based on an exponential HI profile, to estimate the HI mass within the optical radius, and the HI gas surface density (Sect.~2.1). H$_2$ masses (Sect.~4) and total gas surface densities (Sect.~2.1) for the 43 sample XMPs have been derived assuming several empirical molecular mass scaling relations calibrated for samples of dwarf and spiral galaxies.

The XMP HI data (Fig.~1b) approximately follow the relation found for FIGGS galaxies (yellow solid line; Fig.~1b), and are roughly consistent with the position of many of the DGS sources (brown open triangles; Fig.~1a) on the KS plot. However, the HI data for many XMPs already fall close to, or below, the KS law, suggesting that the addition of an H$_2$ component will veer the XMPs towards extremely low (net) star formation efficiencies (Fig.~6b and 6c). Because no direct H$_2$ measurements are available for the XMPs (except for three upper limits), the application of different molecular mass scaling relations, using the metallicity, HI mass, and SFR, provide a wide range of possible H$_2$ masses, anywhere from 0.01 to 5 times the HI mass (Fig.~6a). These H$_2$ masses place the XMPs in the low star formation efficiency regime (SFE$_{\rm gas} \simeq$ 10$^{-9}$ -- 10$^{-12}$ ~yr$^{-1}$; Fig.~6b and 6c). For the lower range of H$_2$ masses (green crosses; Fig.~6a and b), the data may still be consistent with the KMT+ model (black solid line; Fig.~6b) in the HI-dominated regime, but for elevated H$_2$ abundances (red and orange crosses; Fig.~6a -- 6c), this model does not provide a satisfactory interpretation. Further investigation shows that XMPs generally possess large sSFRs (sSFR $\simeq $ 10$^{-10}$ -- 10$^{-7}$~yr$^{-1}$), similar to high redshift star-forming galaxies, but low HI star formation efficiencies (SFE$_{\rm HI} \equiv $1/$\tau _{\rm HI}$ $\simeq $ 10$^{-11}$ --10$^{-8}$~yr$^{-1}$; Fig.~4b). The HI gas depletion timescale is found to be correlated with the HI mass; the larger the HI mass, the longer the HI depletion rate (Fig.~3c). In addition, there is a large variation of the total HI mass present on large-scales (M$_{\rm HI}$/M$_{\odot} \simeq $ 10$^{7}$ --10$^{10}$) for a fixed SFR (Fig.~3d). These findings suggest that the global HI content is likely not an appropriate tracer of the H$_2$ mass nor of the SFR (as given by the KS law).  

The results could be interpreted as the metal-poor HI reservoir (Filho et al. 2013) unavoidably guiding the system towards a lower global star formation efficiency, as in the model developed by Krumholz (2013). Alternatively, the large-scale HI reservoir may not be directly involved in the star formation process. In XMPs, the star formation is clearly dominated by compact star-forming regions which are fed by metal-poor HI clumps, likely accreted via cold cosmological accretion flows (e.g., S\'anchez Almeida et al. 2013, 2014, 2015; Filho et al. 2013, 2015). Star formation triggering timescales (relative to the gas depletion and star formation timescales; Fig.~4b) and galaxy winds (Fig.~5), which are known to be relevant in low mass galaxies such as XMPs, likely also make a significant contribution to the low star formation efficiency in these metal-poor environments. While on small scales, the individual star-forming regions may (e.g., S\'anchez Almeida et al. 2015) or may not (e.g., Shi et al. 2014; Rubio et al. 2015) be efficient at producing stars, the net effect, on large-scales, is the appearance of HI-dominated, high specific star formation rate (sSFR $\gtrsim $ 10$^{-10}$~yr$^{-1}$), inefficient star-forming (SFE$_{\rm gas} \lesssim$ 10$^{-9}$~yr$^{-1}$) systems.

\section{Acknowledgments}

This research has been supported by the {\it Estallidos IV} project (AYA2010-21887-C04-04) and {\it Estallidos V} project (AYA2013-47742-C4-2-P), funded by the Spanish Ministerio de Economia e Competitividad (MINECO). 

We would like to thank D. Hunter for help regarding WLM data, and M. Krumholz for a useful discussion on the KS relation in low metallicity environments.

We thank the anonymous referee for their comments and suggestions, which have helped to improve this manuscript.

This research has made use of the SDSS DR7, the DSS-II, the NED, and the MPA-JHU DR7 release of spectrum measurements.








\begin{appendix}

\subsection{Dicussion on WLM}

WLM has an integrated H${\alpha}$-derived SFR (SFR $\simeq $ 0.002~M$_{\odot}$~yr$^{-1}$; Hunter, Elmegreen \& Ludka 2010) two orders of magnitude above the H${\alpha}$/far-ultraviolet SFR of the individual Northwestern (cloud A) and Southeastern (cloud B) star-forming regions, respectively (SFR$_{\rm NW} \simeq $ 3.9 -- 4.8 $\times$ 10$^{-5}$ and SFR$_{\rm SE} \simeq $ 1.7 -- 12.6 $\times$ 10$^{-5}$~M$_{\odot}$~yr$^{-1}$; Elmegreen et al. 2013). Because the star formation in WLM is likely dominated by a few compact, higher efficient star-forming regions, the integrated SFR surface density (dark green triangle; Fig.~1b), is about an order of magnitude below that of the individual star-forming regions (dark green triangles; Fig.~1a).  

\subsection {Discussion on the XMPs in the DGS Dataset}

The DGS dataset (Madden et al. 2013; Cormier et al. 2014) contains 18 sources in common with the original XMP sample, only three of which possess CO observations that yield H$_2$ mass upper limits: SBS 0335-052, IZw 18 and VII Zw 403 (UGC 6456). SBS 0335-052 is a well-known interacting pair of galaxies (SBS 0335-052 East and SBS 0335-052 West; Izotov et al 1990; Pustilnik et al. 2001), and (individual total) HI mass values (Filho et al. 2013 and references therein) have been adopted, which are consistent with the (total) HI mass quoted by Cormier et al. (2014; log(M$_{\rm HI}$/M$_{\odot}$) $\simeq $ 8.6). A MPA-JHU SFR estimate is not available for this interacting pair. IZw 18 also possesses a companion galaxy (Dufour, Esteban \& Casta\~neda 1996). The total HI mass (log(M$_{\rm HI}$/M$_{\odot}$) $\simeq $ 8.0) adopted by Cormier et al. (2014) for IZw 18 is consistent with the value adopted in the present work (Filho et al. 2013 and references therein). However, the location of IZw 18 on the Cormier et al. (2014) KS plot (Fig.~1a) is over an order of magnitude lower in $\Sigma _{\rm SFR}$ than the estimation in the present work (Fig.~1b), due to the difference in the probed area (4.8~kpc$^2$ versus 1.2~kpc$^2$), and adopted SFR (10$^{-2.3}$~M$_{\odot}$~yr$^{-1}$ versus 10$^{-1.7}$~M$_{\odot}$~yr$^{-1}$), which in Cormier et al. (2014) is sampled in the thermal infrared regime. Although there is no MPA-JHU SFR estimation for VII Zw 403, the adopted total HI mass (log(M$_{\rm HI}$/M$_{\odot}$) $\simeq $ 6.7; Filho et al. 2013 and references therein) is approximately an order of magnitude less than the value given in Cormier et al. (2014; log(M$_{\rm HI}$/M$_{\odot}$) $\simeq $ 7.5). Because it appears that the adopted HI data are the same (Thuan, Hibbard \& L\'evrier 2004), the difference in HI mass is due to different adopted distances (4.5~Mpc in Madden et al. 2013 versus 1.42~Mpc in Filho et al. 2013).

\subsection {Discussion on the XMPs Sextans A and ESO 146-G14}

The adopted total HI masses of log(M$_{\rm HI}$/M$_{\odot}$) $\simeq 7.8 $ and 9.0 for Sextans A and ESO 146-G14, respectively (Filho et al. 2013 and references therein), are similar to the values quoted by Shi et al. (2014) for the disks of these two sources. However, the HI surface density is about an order of magnitude lower than the values provided for the individual star-forming regions (Shi et al. 2014). Similarly, the total HI mass for WLM (log(M$_{\rm HI}$/M$_{\odot}$) $\simeq $ 7.8; Hunter, Elmegreen \& Ludka 2010) is about two orders of magnitude above the estimated HI mass for the Northwestern component (Elmegreen et al. 2013). By averaging over the whole galaxy, although large-scale HI gas and HI gas from diffuse inter-star-forming regions are included, a significant part of the HI gas may be clumpy and concentrated in the star-forming regions. Hence, lower (by approximately an order of magnitude) integrated $\Sigma _{\rm HI}$ values (red open square and circle, and dark green open triangle, respectively; Fig.~1b) are obtained relative to the individual star-forming region values (red open squares and circles, and dark green open triangle, respectively; Fig.~1a). 

The dust-derived total gas surface densities for Sextans A and ESO 146-G14 are high (Shi et al. 2014). Indeed, the disks of Sextans A and ESO 146-G14 (M$_{\rm dust}$/M$_{\odot}$ $\simeq $ 10$^4$ and 6 $\times$ 10$^5$, respectively; Shi et al. 2014) and individual star-forming regions (M$_{\rm dust}$/M$_{\odot}$ $\simeq $ 1 -- 2 $\times$ 10$^3$ and 8 $\times$ 10$^4$ -- 3 $\times$ 10$^5$, respectively; Shi et al. 2014) contain significant dust masses, that exceed the total dust mass (M$_{\rm dust}$/M$_{\odot}$ $\simeq $ 450 -- 1800) determined in the prototypical XMP, IZw 18 (Fisher et al. 2014; Hunt et al. 2014).

It is also noteworthy that the infrared/far-ultraviolet-derived SFRs (on star-forming region scales) quoted for Sextans A (SFR $\simeq $ 10$^{-4.3}$ -- 10$^{-3.1}$~M$_{\odot}$~yr$^{-1}$; Shi et al. 2014) and ESO 146-G14 (SFR $\simeq $ 10$^{-3.1}$ -- 10$^{-2.7}$~M$_{\odot}$~yr$^{-1}$; Shi et al. 2014) are low compared to the optical-derived peak SFRs found for the star-forming regions in a small sample of XMPs (S\'anchez Almeida et al. 2015), and the ultraviolet-derived peak SFRs quoted for Sextans A in Dohm-Palmer et al. (2002). Because the star formation in these XMPs is highly inhomogeneous, the integrated SFR surface density (red open square and circle, respectively; Fig.~1b) is found to be an order of magnitude above that estimated for the individual star-forming regions (red squares and circles, respectively; Fig.~1a).

\end{appendix}
 

\begin{references}

\reference{}Abazajian, K. N., Adelman-McCarthy, J. K., Ag\"ueros, M. A. et al. 2009, ApJS, 182, 543
\reference{}Adamo, A., \"Ostlin, G. \& Zackrisson, E. 2011, MNRAS, 417, 1904
\reference{}Amor\'\i n, R., Mu\~noz-Tu\~n\'on, C., Aguerri, J. A. L. \& Planesas, P. 2015, A\&A, arXiv:1512.06153
\reference{}Bergvall, N. \& R\"onnback, J. 1995, MNRAS, 273, 603
\reference{}Bigiel, F., Leroy, A., Walter, F. et al. 2008, AJ, 136, 2846
\reference{}Bigiel, F., Leroy, A., Walter, F. et al. 2010, AJ, 140, 1194
\reference{}Birnboim, Y. \& Dekel, A. 2003, MNRAS, 345, 349
\reference{}Bolatto, A. D., Leroy, A. K. Jameson, K. et al. 2011, ApJ, 741, 12 
\reference{}Bolatto, A. D., Wolfire, M. \& Leroy, A. K. 2013, ARA\&A, 51, 207
\reference{}Brinchmann, J., Charlot, S., White, S. D. M. et al. 2004, MNRAS, 351, 1151
\reference{}Brooks, A. M., Governato, F., Quinn, T., Brooks, C. B. \& Wadsley, J. 2009, ApJ, 694, 396
\reference{}Ceverino, D., S\'anchez Almeida, J., Mu\~noz-Tu\~n\'on, C. et al. 2015, MNRAS, arXiv:1509.02051
\reference{}Chomiuk, L. \& Povich, M. S. 2011, AJ, 142, 197
\reference{}Cormier, D., Madden, S. C., Lebouteiller, V. et al. 2014, A\&A, 564, 121
\reference{}Cormier, D., Madden, S. C., Lebouteiller, V. et al. 2015, A\&A, 578, 53 
\reference{}Daddi, E. et al. 2010, ApJ, 714, 118
\reference{}Dufour, R. G., Esteban, C. \& Casta\~neda, H. O. 1996, ApJ, 471, L87
\reference{}Dayal, P., Ferrara, A. \& Dunlop, J. S. 2013, MNRAs, 430, 2891
\reference{}da Silva, R. L., Fumagalli, M. \& Krumholz, M. R. 2014, MNRAS, 444, 3275
\reference{}De Young, D. S. \& Gallagher, J. S. 1990, AJ, 356, 15
\reference{}Dekel, A. \& Birnboim, Y. 2006, MNRAS, 368, 2
\reference{}Dohm-Palmer, R. C., Skillman, E. D., Mateo, M. et al. 2002, AJ, 123, 813 
\reference{}Eder, J. \& Schombert, J. M. 2000, ApJS, 131, 47
\reference{}Ekta, B. \& Chengalur, J. N. 2010, MNRAS, 406, 1238
\reference{}Elbaz, D., Daddi, E., Le Borgne, D. et al. 2007, A\&A, 468, 33
\reference{}Elmegreen, B. G., Rubio, M., Hunter, D. A.	et al. 2013, Nature, 495, 487
\reference{}Elmegreen, B. G. \& Hunter, D. A. 2015, ApJ, 805, 145
\reference{}Elmegreen, B. G. 2015, ApJ, 814, 30
\reference{}Filho, M. E., Winkel, B., S\'anchez Almeida, J. et al. 2013, A\&A, 558, 18
\reference{}Filho, M. E., S\'anchez Almeida, J., Mu\~noz-Tu\~n\'on, C. et al. 2015, ApJ, 802, 82
\reference{}Fisher, D. B., Bolatto, A. D., Herrera-Camus, R. et al. 2014, Nature, 505, 186
\reference{}Glover, S. C. O. \& Clark, P. C. 2012a, MNRAS, 421, 9
\reference{}Glover, S. C. O. \& Clark, P. C. 2012b, MNRAS, 426, 377
\reference{}Fumagalli, M., Krumholz, M. R. \& Hunt, L. K. 2010, ApJ, 722, 919
\reference{}Haynes, M. P., Giovanelli, R., Martin, A. M.  et al. 2011, AJ, 142, 170 
\reference{}Huchtmeier, W. K., Gropal, K. \& Petrosian, A. 2005, A\&A, 434, 887
\reference{}Huchtmeier, W. K., Petrosian, A., Gropal, K. \& Kunth, D. 2007, A\&A, 462, 919 
\reference{}Hunt, L. K., Testi, L., Casasola, V. et al. 2014, A\&A, 561, 49
\reference{}Hunt, L. K., Garcia-Burillo, S., Casasola, V. et al. 2015, A\&A, 583, 114
\reference{}Hunter, D. A., Elmegreen, B. G. \& Ludka, B. C. 2010, AJ, 139, 447
\reference{}Izotov,  Y. I., Lipovetsky, V. A., Guseva, N. G., Kniazev, A. Y., \& Stepanian, J. A. 1990, Nature, 343, 238
\reference{}James, B. L., Koposov, S., Stark, D. P., Belokurov, V., Pettini, M. \& Olszewski, E. W. 2015, MNRAS, 448, 2687
\reference{}Kalberla, P. M. W. \& Kerp, J. 2009, ARA\&A, 47, 27
\reference{}Kepley, A. A., Wilcots, E. M., Hunter, D. A. \& Nordgren, T. 2007, AJ, 133, 2242
\reference{}Kennicutt, R. C. 1989, ApJ, 344, 685
\reference{}Kennicutt, R. C. 1998, ApJ, 498, 541
\reference{}Kennicutt, R. C. \& Evans, N. J. 2012, ARA\&A, 50, 531
\reference{}Kent, B. R., Giovanelli, R., Haynes, M. P. et al. 2008, AJ, 136, 713
\reference{}Kere\v {s}, D. Katz, N., Weinberg, D. H., Dav\'e, R. 2005, MNRAS, 363, 2
\reference{}Kova\v{c}, K., Oosterloo, T. A. \& van der Hulst, J. M. 2009, MNRAS, 400, 743
\reference{}Kreckel, K., Platen, E., Arag\'on-Calvo, M. A. 2011, AJ, 141, 4
\reference{}Krumholz, M. R., McKee, C. F. \& Tumlinson, J. 2008, ApJ, 689, 865
\reference{}Krumholz, M. R., McKee, C. F. \& Tumlinson, J. 2009a, ApJ, 693, 216
\reference{}Krumholz, M. R., McKee, C. F. \& Tumlinson, J. 2009b, ApJ, 699, 850
\reference{}Krumholz 2012, M. R., ApJ, 759, 9
\reference{}Krumholz 2013, M. R., MNRAS, 436, 2747
\reference{}Lee, H., Skillman, E. D., Cannon, J. M., Jackson, D. C.,  Gehrz, R. D. et al. 2006, ApJ, 647, 970
\reference{}Lelli, F., Verheijan, M. \& Fraternali, F. 2014a, A\&A, 566, 71
\reference{}Lelli, F., Verheijan, M. \& Fraternali, F. 2014b, MNRAS, 445, 1694
\reference{}Leroy, A., Bolatto, A. D., Simon, J. D. \& Blitz, L. 2005, ApJ, 625, 763
\reference{}Leroy, A., Walter, F., Brinks, E. et al. 2008, AJ, 136, 2782
\reference{}Leroy, A., Bolatto, A., Gordon, K. et al. 2011, ApJ, 737, 12
\reference{}Madden, S. C. 2000, NewAR, 44, 249 
\reference{}Madden, S. C. R\'emy-Ruyer, A., Galametz, M. et al. 2013, PASP, 125, 600
\reference{}Madden, S. C. R\'emy, A., Galliano, F. et al. 2012, IAUS, 284, 141
\reference{}\"Ostlin, G., Amram, P., Bergvall, N., Masegosa, J., Boulesteix, J. \& M\'arquez, I. 2001, A\&A, 374, 800
\reference{}Morales-Luis, A. B., S\'anchez Almeida, J., Aguerri, J. A. L. \&  Mu\~noz-Tu\~n\'on, C. 2011, ApJ, 743, 77
\reference{}P\'erez-Montero, E. 2014, MNRAS, 441, 2663
\reference{}Pustilnik, S. A., Brinks, E., Thuan, T. X. \& Izotov, Y. I. 2001, AJ, 121, 1413
\reference{}R\'emy-Ruyer, A., Madden, S. C., Galliano, F. et al. 2013, A\&A, 557, 95
\reference{}R\'emy-Ruyer, A., Madden, S. C., Galliano, F. et al. 2014, A\&A, 563, 31
\reference{}Roychowdhury, S., Chengalur, J. N., Kaisin, S. S. et al. 2011, MNRAS, 414, 55
\reference{}Roychowdhury, S., Chengalur, J. N., Kaisin, S. S. et al. 2014, MNRAS, 445, 1329
\reference{}Roychowdhury, S., Huang, M. -L., Kauffmann, G., Wang, J. \& Chengalur, J. N. 2015, MNRAS, 449, 3700
\reference{}Rubio,	M., Elmegreen, B. G., Hunter, D. A. et al. 2015, Nature, 525, 218
\reference{}Salim, S., Rich, R. M., Charlot, S. et al. 2007, ApJS, 173, 267
\reference{}S\'anchez Almeida, J., Mu\~noz-Tu\~n\'on, C., Elmegreen, D. M., Elmegreen, B. G., M\'endez-Abreu, J. 2013, ApJ, 767, 74
\reference{}S\'anchez Almeida, J., Elmegreen, B. G., Mu\~noz-Tu\~n\'on, C. Elmegreen, D. M. 2014, A\&ARv, 22, 71
\reference{}S\'anchez Almeida, J., Elmegreen, B. G., Mu\~noz-Tu\~n\'on, C. et al. 2015, ApJL, 810, 15
\reference{}S\'anchez Almeida, J., P\'erez-Montero, E., Morales-Luis, A. B. et al. 2016, ApJ, arXiv:1601.01631 
\reference{}Schmidt, M. 1959, ApJ, 129, 243
\reference{}Schiminovich, D., Catinella, B., Kaumann, G., et al. 2010, MNRAS, 408, 919
\reference{}Schruba, A., Leroy, A. K., Walter, F. et al. 2011, AJ, 142, 37
\reference{}Schruba, A., Leroy, A. K. Walter, F. et al. 2012, AJ, 143, 138
\reference{}Shetty, R., Glover, S. C., Dullemond, C. P. \& Klessen, R. A. 2011, MNRAS, 412, 1686
\reference{}Shi, Y., Armus, L., Helou, G. et al. 2014, Nature, 514, 335
\reference{}Shi, Y., Wang, J., Zhang, Z. -Y. et al. 2015, ApJ, 804, 11 
\reference{}Silich, S. \& Tenorio-Tagle, G. 2001, ApJ, 91, 2001
\reference{}Skillman, E. D., Salzer, J. J., Berg, D. A. et al. AJ, 146, 3
\reference{}Stierwalt, S. Haynes, M. P., Giovanelli, R. et al. 2009, AJ, 138, 338 
\reference{}Thuan, T. X., Hibbard, J. E. \& L\'evrier, F. et al. 2004, AJ, 128, 617
\reference{}Thuan, T. X., Lipovetsky, V. A., Martin, J. -M. \& Pustilnik, S. A. 1999, A\&AS, 139, 1
\reference{}Walter, F., Brinks, E., de Blok, W. J. G., Bigiel, F., Kennicutt, R. C. Jr., Thornley, M. D. \& Leroy, A. 2008, AJ, 136, 2563
\reference{}Whitaker, K., Franx, M., Leja, J. et al. 2014, ApJ, 795, 104
\reference{}Wolfire, M. G., Hollenbach, D. \& McKee, C. F. 2010, ApJ, 716, 1191
\reference{}Wong, T. Xue, R., Bolatto, A. D. et al. 2013, ApJ, 777, 4
\reference{}Wyder, T. K., Martin, D. C., Barlow, T. A. et al. 2009, ApJ, 696, 1834
\reference{}van Zee, L., Skillman, E. D. \& Salzer, J. J. 1998, 116, 1186
\reference{}van Zee, L., Salzer, J. J. \& Skillman E. D. 2001, AJ, 122, 121

\end{references}
\end{document}